\documentclass[twocolumn,showpacs,aps,pra,amsmath,amssymb]{revtex4-2}

\usepackage{mathrsfs}
\usepackage{amsfonts}
\usepackage{amsmath,amsfonts,amssymb,mathrsfs,bm}
\usepackage{verbatim,psfrag,times,CJK}
\usepackage{graphicx,graphics,color,epsfig}
\usepackage{float}
\DeclareMathOperator\arctanh{arctanh}
\usepackage{flafter}
\usepackage{subeqnarray}
\usepackage{cases}
\usepackage{amsmath, upgreek, amssymb, graphicx, subfigure, paralist, dsfont}
\usepackage[colorlinks, linkcolor=blue, citecolor=blue, urlcolor=blue, breaklinks=true]{hyperref}

\definecolor{red}{rgb}{1,0,0}

\begin{document}

\title{Nonreciprocal flow of fluctuations, populations and correlations between doubly coupled bosonic modes}

\author{Zbigniew Ficek}
\email{z.ficek@if.uz.zgora.pl}

\address{Quantum Optics and Engineering Division, Institute of Physics, University of Zielona G\'ora, Szafrana 4a, 65-516 Zielona G\'ora, Poland}

\date{\today}



\begin{abstract}
Interesting new correlation and unidirectional properties of two bosonic modes under the influence of environment appear when the modes are mutually coupled through the simultaneously applied linear mode-hopping and nonlinear squeezing interactions. Under such double coupling, it is found that while the Hamiltonian of the system is clearly Hermitian the dynamics of the quadrature components of the field operators can be attributed to non-Hermicity of the system. It is manifested in an asymmetric coupling between the quadrature components which then leads to a variety of remarkable features. In particular, we identify how the emerging exceptional point controls the conversion of thermal states of the modes into single-mode classically or quantum squeezed states. Furthermore, for reservoirs being in squeezed states, we find that the two-photon correlations present in these reservoirs are responsible for unidirectional flow of populations and correlations among the modes and the flow can be controlled by appropriate tuning of the mutual orientation of the squeezed noise ellipses. In the course of analyzing these effects we find that the flow of the population creates the first-order coherence between the modes which, on the other hand rules out an enhancement of the two photon correlations responsible for entanglement between the modes. These results suggest new alternatives for the creation of single mode squeezed fields and the potential applications for controlled unidirectional transfer of population and correlations in bosonic chains.
\end{abstract}

\maketitle

\section{Introduction}

It has been demonstrated in recent years that unconventional properties of non-Hermitian parity-time $({\cal PT})-$ symmetric systems such as exceptional points and nonreciprocity~\cite{he12,ma19,or19,am20,bb21,df22} are not restricted to the time reversal systems~\cite{em18,pm23,az23,zz24,pb25,bx25}, but can be obtained in quantum systems with a Hermitian Hamiltonian by interfering linear (excitation-preserving) and nonlinear two-mode interactions~\cite{mt18,dp18,wc19,fc20,fc21,ps22,ws23,sw24,yt25}. Subsequently, similar approaches have been studied including simultaneous application of linear and dissipative interactions~\cite{ba18,am24} or creation of two coupling processes by engineering collective atomic systems~\cite{de07,wf09}. The simultaneous existence of two coupling processes between quantum systems has been experimentally realized among optomechanical cavities~\cite{rm16,sw24} and in superconducting circuits~\cite{sh15,zl19}.

To date, most of the treatments of the dynamics of doubly coupled modes have either been concerned with entanglement dynamics~\cite{vs18,yt25,vc25}, quantum sensing~\cite{mc20,lz22} and topological phases~\cite{hh11,mp15,ph16,ga18,wb20}. Here, we consider a system composed of two Gaussian bosonic modes and concentrate on their fluctuation and correlation properties for features indicative of the simultaneous presence of the linear and nonlinear couplings processes. The two mode system provides the simplest example of the effects generated by the double coupling. In particular, we investigate how the coupling influences on the fluctuations, populations and correlation properties of the modes. The treatment includes the dissipation of the modes to local thermal as well as to squeezed reservoirs which, as we will see can have a significant effect on the fluctuation and correlation properties of the modes. 

Modes interacting with local thermal or squeezed reservoirs will be sensitive to the number of photons and in the case of squeezed reservoirs also to the correlations present in the reservoirs and in general will evolve in a phase-sensitive fashion to a stationary state that will reflect such correlations. A squeezed reservoir is characterized by the mean photon number $n$ and by the phase dependent two-photon correlations $m$, which range from zero to $|m|=\sqrt{n(n+1)}$. Values of the correlation in the range $\sqrt{n(n+1)}\geq |m| >n$ correspond a quantum squeezed reservoir while values of the correlation in the range $0< |m| \leq n$ correspond to a classically squeezed reservoir, and $m=0, n\neq 0$ correspond to a thermal reservoir~\cite{df04,fw14}. Presence of the correlations $m$ leads to a reduction of the fluctuations in one quadrature of the modes.

It is well known that when only linear coupling is applied between two modes, it cannot create any correlations between noisy Gaussian modes~\cite{mw95}. On the other hand when the nonlinear coupling is applied, it can create two-photon correlations between the modes which are necessary for entanglement but the correlated modes are left mutually incoherent~\cite{mg96,lm98,sl12,hm15,mh19,sl22}. Moreover the modes are strongly amplified in both populations and fluctuations such that the modes are found in a highly fluctuating thermal state~\cite{mg67a,mg67b,bk85}.  

In this paper, we demonstrate that when both linear and nonlinear interactions are simultaneously applied the fluctuation and correlation properties of the modes are significantly different. Analytical expressions are obtained for steady state variances of the quadrature components of the mode operators and for correlation functions which show that relative to mutual coupling strengths of the two interactions an exceptional point emerges which is characteristic of non-Hermitian systems. To put it another way, our results demonstrate that in the Hermitian quantum system composed of simultaneously linearly and nonlinearly coupled modes one can construct non-Hermitian dynamics which can lead to nonreciprocal (one directional) influences of the modes on each other. It also creates an exceptional point which separates two distinctive parameter regimes, an exponential amplification regime and an oscillatory regime. In these two regimes the fluctuation properties of the modes are found to be differently altered even to the point of turning the fluctuations from thermal to quantum squeezed fluctuations. Under suitable conditions, the double coupling can establish nonresiprocal transfer of population and correlations between the modes.

The plan of this paper is as follows. In Sec.~\ref{sec2}, we introduce the Hamiltonian of doubly coupled bosonic modes and give a description of independent external reservoirs to which the modes are dumped. Then, we derive the quantum Langevin equations for the quadrature components of the field operators and discuss their nonresiprocal coupling properties. 
In Sec.~\ref{sec3}, we give general solutions for the time-dependent quadrature operators which exhibit the presence of an exceptional point separating two parameter regimes, exponential amplification and oscillatory regimes. In Sec.~\ref{sec4}, we specialize to the exponential amplification regime and apply the solutions to obtain analytic stationary expressions of physical quantities of interest as the variances of the quadrature operators, populations of the modes, single- and two-mode correlation functions, and to investigate their dependence on the noise properties of thermal and squeezed reservoirs. Analytic expressions of these physical quantities in the oscillatory regime are presented and extensively discussed in Sec.~\ref{sec5}. Differences and similarities of the results obtained in those two regimes are also examined. Finally, a brief discussion of the results and conclusions are given in Sec.~\ref{sec6}.

\section{Doubly coupled bosonic modes}\label{sec2}

\begin{figure}[h]
\includegraphics[width=5.0 cm]{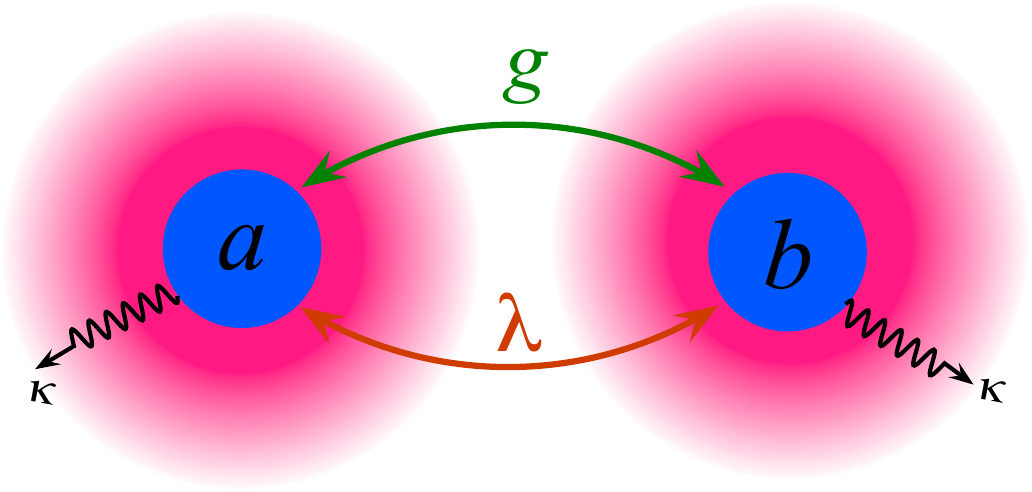}
\caption{Schematic diagram of the considered system composed of two modes $a$ and $b$ coupled to each other through linear $(\lambda)$ and nonlinear $(g)$ interactions. The modes are dumped with rate $\kappa$ by the interaction with local broadband reservoirs. \label{fig1}}
\end{figure}

The system we study consists of two frequency degenerate radiation modes described by bosonic creation (annihilation) operators, $a^{\dag} (a)$ and $b^{\dag} (b)$, respectively. The modes are directly coupled to each other through presence of two different types of interaction processes, the linear optical photon exchange (beamsplitter type) and nonlinear two-photon (two-mode squeezing) interaction processes. In addition, both modes are damped with the rate $\kappa$ by coupling to local (independent) reservoirs, which will be considered as thermal or squeezed reservoirs, see Fig.~\ref{fig1}. 

The interaction between the modes is determined by the Hamiltonian $(\hbar =1)$
\begin{align}
H &= \lambda\left(a^{\dag}be^{-i(\phi_{a}-\phi_{b})} + a b^{\dag}e^{i(\phi_{a}-\phi_{b})}\right) \nonumber\\
&+ g\left(abe^{i(\phi_{a}+\phi_{b})} + a^\dag b^{\dag}e^{-i(\phi_{a}+\phi_{b})}\right)  ,\label{01}
\end{align}
where $\lambda$ and $g$ are real parameters determining the strength, respectively, the linear  and nonlinear coupling between the modes, and $\phi_{i}\, (i=a,b)$ is the phase of the $i$-the mode. The simultaneous action of the two interactions will couple the modes and will give the systematic evolution of the modes due to the mutual interaction. Without loss of generality we will assume that the phase of the mode $a$ is fixed at $\phi_{a}=0$, while the phase $\phi_{b}\equiv \phi$ of the mode $b$ can be varied that $\phi$ will play the role of the relative phase between squeezed reservoirs.

It has been demonstrated that the presence of the nonlinear two-photon interaction process in the interaction Hamiltonian~(\ref{01}) leads to nonreciprocal and non-Hermitian behaviour of the system~\cite{dp18,ps22,ws23,sw24,yt25}. To see this, we introduce Hermitian operators representing the amplitudes of two quadrature phase components of the fields 
\begin{align}
X_{a} &= \frac{1}{\sqrt{2}}\left(a+a^{\dag}\right) ,\qquad Y_{a} =\frac{1}{\sqrt{2}i}\left(a-a^{\dag}\right) ,\nonumber\\
X_{b} &= \frac{1}{\sqrt{2}}\left(be^{i\phi}+b^{\dag}e^{-i\phi}\right) ,\quad Y_{b} =\frac{1}{\sqrt{2}i}\left(be^{i\phi} -b^{\dag}e^{-i\phi}\right) ,\label{02}
\end{align}
and find that the quadrature componets$X_{i}$ and $Y_{j}$, $(i\neq j=a,b)$ quadrature components satisfy a set of coupled Heisenberg equations of motion
\begin{align}
i\frac{d}{dt}\Phi(t) = \left[\Phi(t),H\right] = {\cal A}\Phi(t) ,\label{03}
\end{align}
where $\Phi(t)=[X_{i}(t),Y_{j}(t)]^{T}$ and 
\begin{align}
{\cal A} =  \left(
\begin{array}{cc}
0 & -i(g-\lambda)  \\
-i(g+\lambda) & 0
\end{array}
\right) \label{04}
\end{align}
is the evolution matrix.

It is easily verified that the matrix ${\cal A}$ is non-Hermitian when $g\neq 0$. Thus, the presence of the nonlinear process in the interaction between the modes induces non-Hermitian dynamics in the system even though the Hamiltonian is Hermitian. 

Assuming further that the modes interact with local reservoirs, which results in damping of the modes with a rate $\kappa$ and quantum noise of the reservoir incident on the modes, the evolution of the quadrature components obeys the quantum Langevin equation 
\begin{align}
\frac{d}{dt}\Phi(t) = -{\cal M}\Phi(t) -\sqrt{2\kappa} \Phi^{\rm in}(t) ,\label{05}
\end{align}
where $\Phi^{\rm in}(t)=[X^{\rm in}_{i}(t),Y^{\rm in}_{j}(t)]^{T}$, and
\begin{align}
{\cal M} =  \left(
\begin{array}{cc}
\kappa & (g-\lambda)  \\
(g+\lambda) & \kappa
\end{array}
\right) .\label{06}
\end{align}
We first observe that inside each pair of equations, quadratures are coupled to each other in an asymmetric manner that the $X$ quadratures are coupled to $Y$ quadratures with strength $(g-\lambda)$, whereas $Y$ quadratures are coupled to $X$ with strength $(g+\lambda)$. For $g = \lambda$ we have complete coupling asymmetry that the evolution of the $X$ quadratures is decoupled from the evolution of the $Y$ quadratures. In this case there is a unidirectional coupling between the quadratures. Thus, we expect chiral propagation of fluctuations and correlations from $Y$ quadratures to $X$ quadratures of the other mode. 

The dynamics of the quadrature operators is influenced by the quantum noise operators $X_{i}^{\rm in}$ and $Y_{i}^{\rm in}$ of the input modes to which the modes $a$ and $b$ are coupled. They obey Gaussian statistics and are delta correlated in time. We assume that the input noise operators represent two independent reservoirs which could be in squeezed vacuum states. Typical sources of a squeezed vacuum field are degenerate parametric oscillators (DPO) operating below threshold, whose the fields filling the input modes results in the input correlation functions~\cite{dw83,cg84,lk87,zz90,df99,gz00}
\begin{align}
\langle X_{i}^{{\rm in}}(t)X_{i}^{{\rm in}}(t^{\prime})\rangle &= \frac{1}{2}\left(- \frac{\nu^{2}-\mu^{2}}{2\nu}e^{-\nu|t-t^{\prime}|} + \delta(t-t^{\prime}) \right) ,\nonumber \\
\langle Y_{i}^{{\rm in}}(t)Y_{i}^{{\rm in}}(t^{\prime})\rangle &=  \frac{1}{2}\left(\frac{\nu^{2}-\mu^{2}}{2\mu}e^{-\mu|t-t^{\prime}|} + \delta(t-t^{\prime}) \right),\label{07}
\end{align}
where $\mu=\frac{1}{2}\gamma_{c} -\varepsilon, \nu=\frac{1}{2}\gamma_{c}+\varepsilon$, in which $\gamma_{c}$ is the damping rate of the DPO cavity and $\varepsilon$ is its amplification amplitude, proportional to the amplitude of a classical field pumping the DPO cavity. The parameters $\mu$ and $\nu$ determine the bandwidth of the squeezed reservoirs. 

In the case when $\mu$ and $\nu$ are much larger than $\kappa$, corresponding to broadband reservoirs, the exponentials appearing in Eq.~(\ref{07}) can be approximated by $\delta$ functions and then the correlations functions (\ref{07}) take the form 
\begin{align}
\langle X_{a}^{{\rm in}}(t)X_{a}^{{\rm in}}(t^{\prime})\rangle &= \left(\frac{1}{2}+n+m\right)\delta(t-t^{\prime}) ,\nonumber \\
\langle Y_{a}^{{\rm in}}(t)Y_{a}^{{\rm in}}(t^{\prime})\rangle &= \left(\frac{1}{2}+n-m\right)\delta(t-t^{\prime}) ,\nonumber \\
\langle X_{b}^{{\rm in}}(t)X_{b}^{{\rm in}}(t^{\prime})\rangle &= \left(\frac{1}{2}+n+m\cos2\phi\right)\delta(t-t^{\prime}) ,\nonumber \\
\langle Y_{b}^{{\rm in}}(t)Y_{b}^{{\rm in}}(t^{\prime})\rangle &= \left(\frac{1}{2}+n-m\cos2\phi\right)\delta(t-t^{\prime}) ,\label{08}
\end{align}
where
\begin{align}
n=\frac{\nu^{2}-\mu^{2}}{4}\left(\frac{1}{\mu^{2}} -\frac{1}{\nu^{2}}\right) ,\quad  m=\frac{\nu^{2}-\mu^{2}}{4}\left(\frac{1}{\mu^{2}} +\frac{1}{\nu^{2}}\right) ,\label{09}
\end{align}
from which we see that by varying the parameters $\gamma_{c}$ and $\varepsilon$ of the parametric oscillator one can produce squeezed vacuum field of a desired values of $n$ and $m$. It is convenient to present results in terms of $n$ and $m$ rather than $\gamma_{c}$ and $\varepsilon$ since 
$n$ describes the average number of thermal photons and $m$ describes the degree of two-photon correlations between photons contained in the reservoirs. It is see from Eq.~(\ref{08}) that the presence of the two-photon correlations $m$ leads to an asymmetric distribution of noise between the quadratures. We assume that the phase of the squeezed reservoir coupled to the mode $a$ is fixed at zero, whereas the phase $\phi$ of the squeezed reservoir coupled to the mode $b$ can be varied. The parameters $n$ and $m$ are related to each other such that $m\leq n$ corresponds to a classically squeezed field whereas $n< m\leq \sqrt{n(n+1)}$ corresponds to a quantum squeezed field~\cite{df04,fw14}.

\section{Dynamics of the doubly coupled bosonic modes}\label{sec3}

To proceed with the solution of the set of differential equations~(\ref{04}), it is convenient to introduce the Laplace transforms of the quadratures
\begin{align}
X_{i}(p) \equiv\int_{0}^{\infty} X_{i}e^{-pt}dt  , \quad Y_{i}(p)\equiv \int_{0}^{\infty} Y_{i}e^{-pt}dt  ,\label{06}
\end{align}
whose the application converts the equations into a set of algebraic equations. The set of the algebraic equations can be readily solved for the transformed quadratures to give 
\begin{align}
X_{i}(p) &= \frac{(p+\kappa)L_{ix}(p) -(g-\lambda)L_{jy}(p)}{(p-p_{1})(p-p_{2})} ,\nonumber\\
\nonumber\\
Y_{i}(p) &= \frac{(p+\kappa)L_{iy}(p) -(g+\lambda)L_{jx}(p)}{(p-p_{1})(p-p_{2})} , \quad i\neq j =a,b ,\label{07}
\end{align}
in which we have introduced the abbreviations
\begin{align}
L_{ix}(p) &= \left[X_{i}(0) -\sqrt{2\kappa}X_{i}^{\rm in}(p)\right] ,\nonumber\\
 L_{iy}(p) &= \left[Y_{i}(0) -\sqrt{2\kappa}Y_{i}^{\rm in}(p)\right] ,\label{08}
\end{align}
and $p_{1}$ and $p_{2}$ are the roots of the quadratic equation
\begin{align}
p^{2}+2\kappa p +\kappa^{2}+\lambda^{2}-g^{2} .\label{09}
\end{align}
The roots can be easily computed, to give
\begin{align}
p_{1,2} = -\kappa \pm \sqrt{g^{2} - \lambda^{2}} . \label{10}
\end{align}
It is clearly seen from Eq.~(\ref{10}) that the roots are strongly dependent on the relationship between the coupling constants $g$ and $\lambda$. We distinguish two parameter regimes that
depending on whether $\lambda >g$ or $\lambda <g$ the factor $\sqrt{g^{2} - \lambda^{2}}$ can be either the real or the complex parameter. A threshold at which the roots change character is for $g=\lambda$, which corresponds to the exceptional point. For $g>\lambda$ the roots are real
\begin{align}
p_{1}= -\kappa +\alpha  ,\quad p_{2} = -\kappa -\alpha  ,\quad \alpha =\sqrt{g^{2}-\lambda^{2}} ,\label{11}
\end{align}
while for $\lambda>g$ the roots are complex 
 \begin{align}
p_{3}= -\kappa +i\beta ,\quad p_{4} = -\kappa -i\beta , \quad \beta= \sqrt{\lambda^{2}-g^{2}} .\label{12}
\end{align}
Thus there are two distinct regimes in which the solutions have different character. We will call the regime $g>\lambda$ an exponential amplification regime, and $\lambda > g$ an oscillatory regime. 

For each of the regimes we will investigate properties of the steady state $(t\rightarrow \infty)$ variances of the quadrature operators, populations of the modes, $\langle a^{\dag}a\rangle$ and $\langle b^{\dag}b\rangle$, single-mode two-photon correlations, $\langle a^{2}\rangle$ and $\langle b^{2}\rangle$, which we will call as "local two-photon correlations", and two-mode two-photon correlations $\langle ab\rangle$,~\cite{ga86,lp16} which we will call as "global two-photon correlations". 
Apart from the global two-photon correlations we will also consider the single-photon two-mode correlations $\langle a^{\dag}b\rangle$, which are known to carry information about coherence properties the modes~\cite{mw95}. This restricted class of the correlation functions considered here results from the fact that the dynamics of the modes considered are Gaussian.
Note that the local two-photon correlations are responsible for squeezing of the single-mode fluctuations, whereas the global two-photon correlations are necessary for entanglement~\cite{hh96,rs00,dg00,gm03,sw05,hz06,ad10}.

By inverse Laplace transformation and application of the Cauchy residue theorem we then have from Eq.~(\ref{07}), with the help of Eq.~(\ref{11}) that in the case of $g>\lambda$ the time evolution of the quadrature operators is given by
\begin{align}
X_{i}(t) = &\frac{1}{2}\left\{[L_{ix}(p_{1})-uL_{jy}(p_{1})]e^{p_{1}t}\right. \nonumber\\
&\left. \qquad \qquad +[L_{ix}(p_{2})+uL_{jy}(p_{2})]e^{p_{2}t} \right\} ,\nonumber\\
Y_{i}(t) &= \frac{1}{2u}\left\{-[L_{jx}(p_{1})-uL_{iy}(p_{1})]e^{p_{1}t}\right. \nonumber\\
&\left. +[L_{jx}(p_{2})+uL_{iy}(p_{2})]e^{p_{2}t} \right\} ,\ i\neq j=a,b .\label{13}
\end{align}
where $u =\sqrt{(g-\lambda)/(g+\lambda)}$.

Similarly, the inverse Laplace transformation of Eq.~(\ref{07}) with the complex roots (\ref{12}) yields to
\begin{align}
X_{i}(t) &= \frac{1}{2}\left\{[L_{ix}(p_{3})-iwL_{jy}(p_{3})]e^{p_{3}t}\right. \nonumber\\
&\left.  +[L_{ix}(p_{4})+iwL_{jy}(p_{4})]e^{p_{4}t} \right\} ,\nonumber\\
Y_{i}(t) &= \frac{i}{2w}\left\{[L_{jx}(p_{3})-iwL_{iy}(p_{3})]e^{p_{3}t}\right. \nonumber\\
&\left.  -[L_{jx}(p_{4})+iwL_{iy}(p_{4})]e^{p_{4}t} \right\} ,\quad i\neq j=a,b .\label{14}
\end{align}
where $w =\sqrt{(\lambda-g)/(\lambda+g)}$. In this case the quadrature evolve in an oscillatory manner.

The dynamics of the quadrature operators is influenced by the quadratures $X_{i}^{\rm in}(p)$ and $Y_{i}^{\rm in}(p)$ of the input modes, whose the statistics are given by the statistics of  the reservoirs to which the modes are coupled. The effect of the statistics of the reservoirs will be evident in the fluctuation and correlation properties of the modes.

\section{Exponential amplification regime, $g>\lambda$.}\label{sec4}

We begin our discussion of the correlation and fluctuation properties of the modes by considering first the exponential regime $g>\lambda$. 

Let us first consider the average number of photons of the modes, i.e. populations of the modes. Since
\begin{align}
a &= \frac{1}{\sqrt{2}}\left(X_{a}+iY_{a}\right) ,\qquad  a^{\dag} = \frac{1}{\sqrt{2}}\left(X_{a}-iY_{a}\right) ,\nonumber\\
b &= \frac{1}{\sqrt{2}}\left(X_{b}+iY_{b}\right) ,\qquad  b^{\dag} = \frac{1}{\sqrt{2}}\left(X_{b}-iY_{b}\right) ,\label{15}
\end{align}
and we have solutions for the time evolution of the quadrature operators, the populations can be evaluated from 
\begin{align}
\langle a^{\dag}a\rangle &= \frac{1}{2}\left(\langle X^{2}_{a}\rangle +\langle Y^{2}_{a}\rangle -1\right) ,\nonumber\\
\langle b^{\dag}b\rangle &= \frac{1}{2}\left(\langle X^{2}_{b}\rangle +\langle Y^{2}_{b}\rangle -1\right) ,\label{16}
\end{align}
where $\langle X_{i}^{2}\rangle$ and $\langle Y_{i}^{2}\rangle$ are variances of the quadrature operators of the modes. Thus, evaluation of the populations of the modes requires calculations of the variances of the quadrature operators of the modes. 

The Gaussian character of the system enables the quadrature variances to be readily calculated. With the use of Eqs.~(\ref{13}) and (\ref{05}), we find that the variances of the quadrature components of the mutually doubly coupled modes are
\begin{align}
\langle X_{a}^{2}\rangle &= \left(\frac{1}{2}+n+m\right)\left(1 +\frac{1}{2}\sinh^{2}\psi\right)\nonumber\\
& +\frac{1}{2}\left(\bar{g}-\bar{\lambda}\right)^{2}\left(\frac{1}{2}+n-m\cos2\phi\right)\cosh^{2}\psi ,\nonumber\\
\langle Y_{a}^{2}\rangle &= \left(\frac{1}{2}+n-m\right)\left(1 +\frac{1}{2}\sinh^{2}\psi\right) \nonumber\\
&+ \frac{1}{2}\left(\bar{g}+\bar{\lambda}\right)^{2}\left(\frac{1}{2}+n+m\cos2\phi\right)\cosh^{2}\psi  ,\nonumber\\
\langle X_{b}^{2}\rangle &= \left(\frac{1}{2}+n+m\cos2\phi\right)\left(1 +\frac{1}{2}\sinh^{2}\psi\right)\nonumber\\
& +\frac{1}{2}\left(\bar{g}-\bar{\lambda}\right)^{2}\left(\frac{1}{2}+n-m\right)\cosh^{2}\psi ,\nonumber\\
\langle Y_{b}^{2}\rangle &= \left(\frac{1}{2}+n-m\cos2\phi\right)\left(1 +\frac{1}{2}\sinh^{2}\psi\right) \nonumber\\
&+\frac{1}{2}\left(\bar{g}+\bar{\lambda}\right)^{2}\left(\frac{1}{2}+n+m\right) \cosh^{2}\psi .\label{17}
\end{align}
where all system parameters have been normalised by the damping rate $\kappa$: $\bar{g}=g/\kappa$, $\bar{\lambda}=\lambda/\kappa$, $\bar{\alpha}=\alpha/\kappa$, and the angle $\psi$ is defined through the relation $\psi = \arctanh(\bar{\alpha})$, $\psi\in (0,\infty)$.

Since $\tanh\psi\leq 1$, we see that expressing the solutions (\ref{17}) in terms of the angle $\psi$ constraints $\bar{\alpha}$ to values $\bar{\alpha}<1$ (threshold occurs at $\bar{\alpha}=1$) at which the solutions are stable. From this constraint we have that also $\bar{g}$ and $\bar{\lambda}$ are constrained to $\bar{\lambda}\leq \bar{g}<1$. 
In what follows we will consider the population and correlation properties of the system in that constrained range of the parameters. Note that there is no constraint on $n$, i.e. the solutions (\ref{17}) are valid for $n\in(0,\infty)$. For the clarity of the expressions we will drop the bar on $\bar{g}, \bar{\lambda}$, and $\bar{\alpha}$ with the understanding that we are dealing with dimensionless rescaled quantities.

Equation (\ref{17}) shows that the double coupling between the modes enhances the variances indicating enhancement of the fluctuations of the modes. However, the variances are unequally  enhanced that the variances of the $X$ quadratures are less enhanced than the variances of the $Y$ quadratures. This is easily understood since, according to Eq.~(\ref{04})  there is an asymmetry in the coupling between the $X$ and $Y$ quadratures leading to unbalanced transfer of fluctuations between the quadratures.

Once the variances have been determined, it is only a matter of substitution of Eq.~(\ref{17}) in Eq.~(\ref{16}) to derive explicit expressions for the steady state populations. Thus, we obtain the following expressions for the populations
\begin{align}
\langle a^{\dag}a\rangle &=  n +\left[\left(\frac{1}{2}+n\right)g^{2}+g\lambda m\cos2\phi \right]\cosh^{2}\psi ,\label{18}
\end{align}
and
\begin{align}
\langle b^{\dag}b\rangle &= n +\left[\left(\frac{1}{2}+n\right)g^{2}+g\lambda m\right]\cosh^{2}\psi .\label{19}
\end{align}
The first terms on the right-hand sides of Eqs.~(\ref{18}) and (\ref{19}) represent the population of the modes in the absence of the direct coupling between the modes. The second terms on the right-hand sides are due to the coupling between the modes. Of particular interest to us is the dependence of the populations on the term $g\lambda$, which accounts for double coupling between the modes. 

It is quite evident from Eqs.~(\ref{18}) and (\ref{19}) that the populations are amplified by the mutual couplings and the major role in the amplification plays the nonlinear coupling process. Since $\tanh\psi$ is limited to one, it means that the amplifier is operating below threshold for the stable steady-state solutions~\cite{mg67a,mg67b,cg84}. 

Note that the manner in which the populations are amplified by the couplings is different for $\langle a^{\dag}a\rangle$ and $\langle b^{\dag}b\rangle$, and the populations are very dependent on the nature of the reservoirs. Perhaps the most significant is the insensitivity of the population of the mode $b$ to the phase of the squeezed reservoir to which the mode is coupled. As it is seen from Eq.~(\ref{18}) the phase sensitivity is transferred to the mode $a$. Thus, a variation of the phase $\phi$ of the squeezed reservoir coupled to the mode $b$ will modify the population of the other mode. 

The squeezed reservoirs provide the mechanism for controlled amplification of the mode $a$ that the population of the mode can vary with the phase in such a way that it could be possible to cease the amplification process. Specifically, at the exceptional point, when $\lambda=g$ with $m=\sqrt{n(n+1)}$ and in the strong squeezing limit $n\gg 1$,  Eq.~(\ref{18}) takes a simple form
\begin{align}
\langle a^{\dag}a\rangle &=  n +\left(\frac{1}{2}+n\right)\left(1+\cos2\phi \right)g^{2} .\label{20}
\end{align}
This result directly shows that for strongly squeezed reservoirs the population is extremely phase sensitive at the exceptional point, and the choice of phase $\phi=\pi/2$ leads to ceasing of the amplification process.

It is worth noting that the phenomenon of ceasing of the amplification process of mode $a$, derived using $m=\sqrt{n(n+1)}$, is unique to quantum squeezing that it is not possible to cease the the amplification process when $m=n$, i.e. the modes interact with classically squeezed fields of the reservoirs. In addition, this phase-dependent amplification of the mode gives us a new control over nonlinear effects in the system of coupled modes. 
\begin{figure}[H]
\includegraphics[width=7.0 cm]{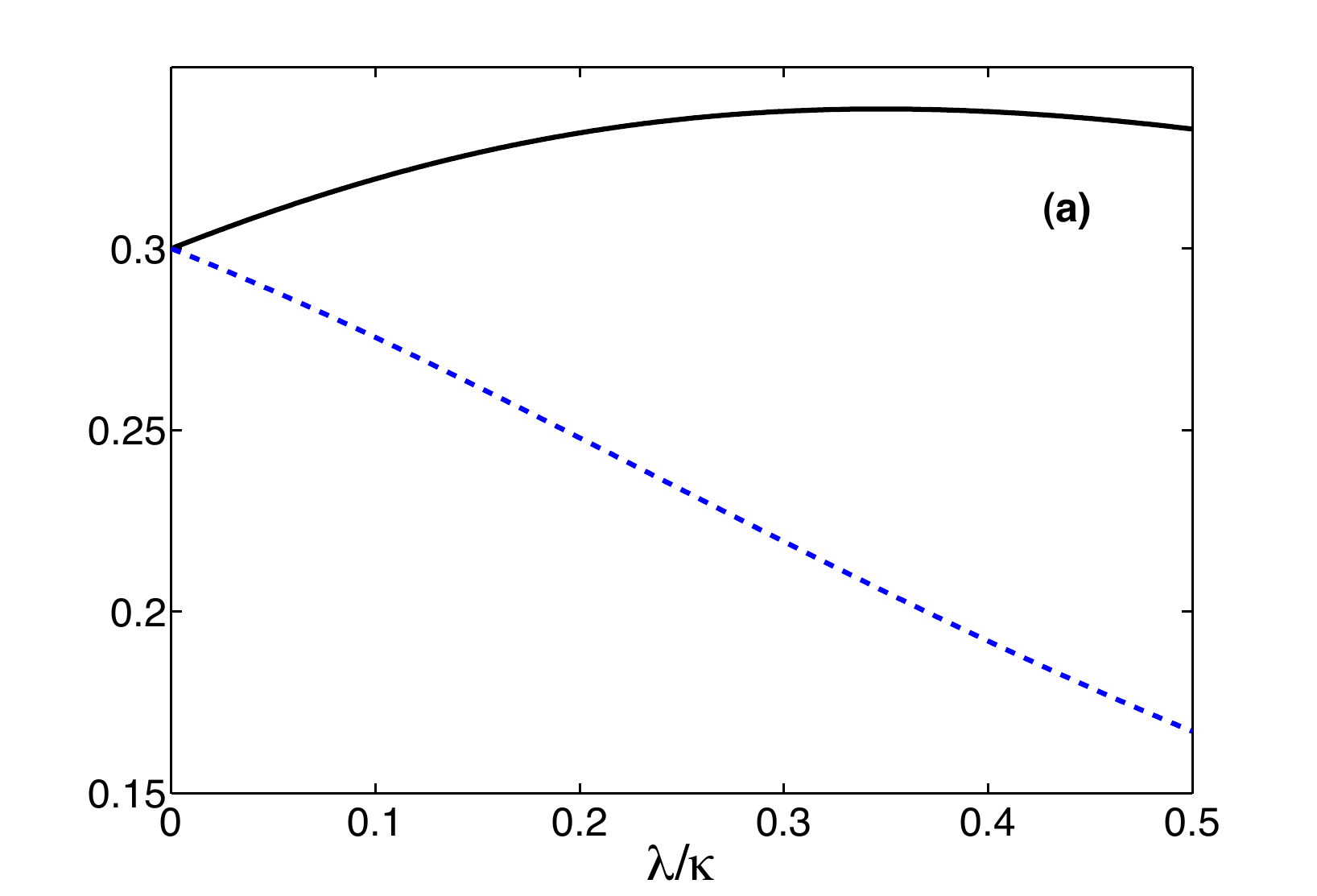}\\
\includegraphics[width=7.0 cm]{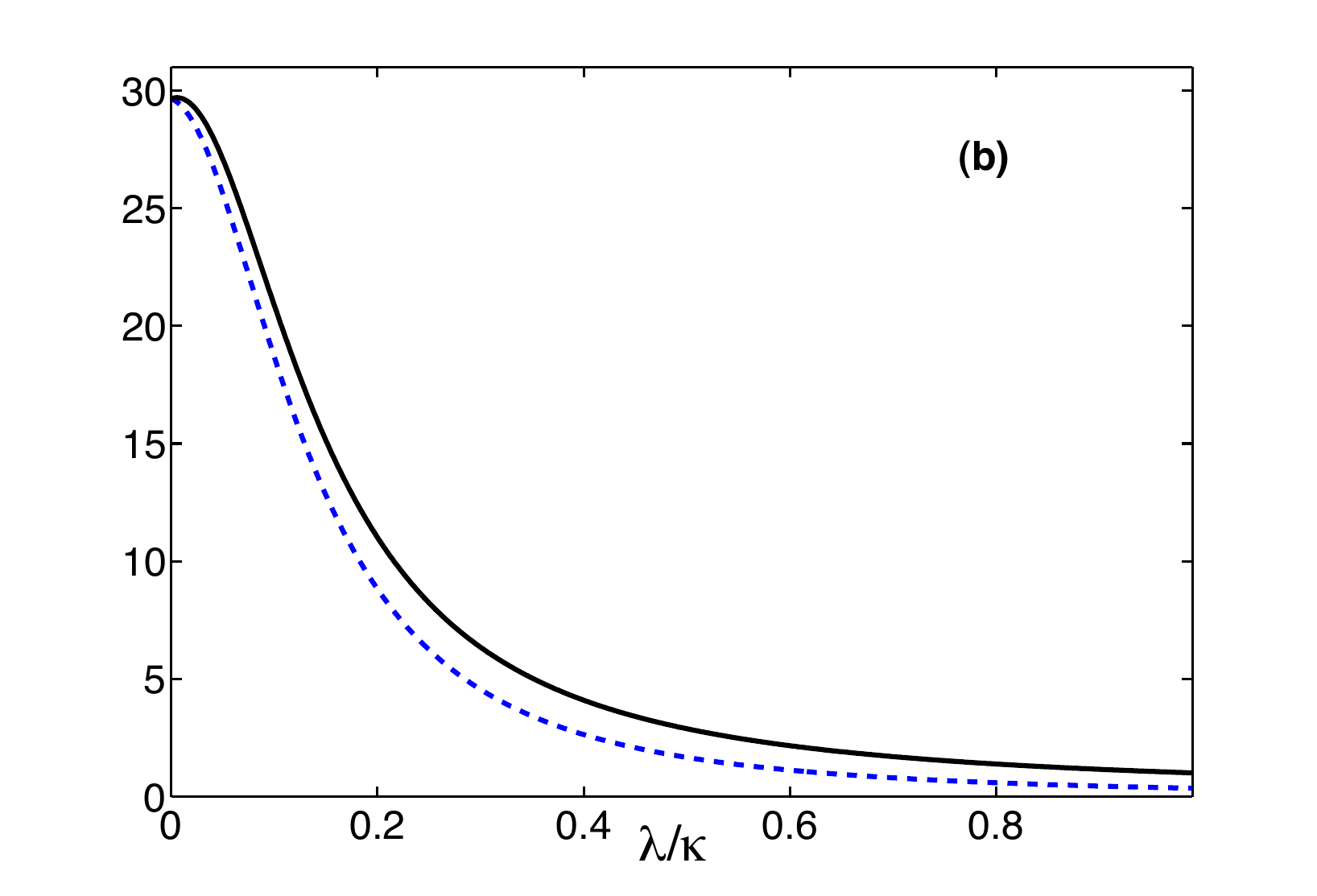}
\caption{Variation of the population of the mode $a$ with  the strength of the linear coupling $\lambda/\kappa$ starting from the purely nonlinear coupling $\lambda=0$ to the exceptional point $\lambda=g$, for $n=0.1$, $m=\sqrt{n(n+1)}$ and two different values of $g$: (\textbf{a}) $g =0.5\kappa$ and (\textbf{b}) $g=0.99\kappa$. The solid black line is for $\phi =0$, the dashed blue line is for $\phi=\pi/2$. \label{fig2}}
\end{figure}

Figure~\ref{fig2} shows the variation of the population of the mode $a$ with the strength of the linear coupling $\lambda$ for two limiting cases of the phase $\phi=0$ and $\phi=\pi/2$ and two different values of the nonlinear coupling strength $g$. Note that the limit $g\rightarrow 1$ corresponds to approaching the instability threshold in the nonlinear amplifying interaction~\cite{mg67a,mg67b,cg84}. For $g$ away from the instability threshold the population varies with the phase and the linear coupling such that for $\phi=0$ it is  further amplified and de-amplified for $\phi=\pi/2$. For $g$ close to the instability threshold, the population exhibits a weak dependence on the phase, rapidly degrades with increasing strength of the linear interactions and returns to its initial value as $\lambda$ approaches the exceptional point $\lambda=g$.

\subsection{Two-photon correlations inside the modes}

We now turn to the calculation of the single mode two-photon correlation functions, which are known to give rise to squeezing of the fluctuations of the quadrature operators.
To evaluate the correlation functions we use Eq.~(\ref{15}) and find
\begin{align}
\left\langle aa\right\rangle &= \frac{1}{2}\left\{\left\langle X^{2}_{a}\right\rangle -\left\langle Y^{2}_{a}\right\rangle +2i\left\langle X_{a}Y_{a}\right\rangle +1\right\} ,\nonumber\\
\left\langle bb\right\rangle &= \frac{1}{2}\left\{\left\langle X^{2}_{b}\right\rangle -\left\langle Y^{2}_{b}\right\rangle +2i\left\langle X_{b}Y_{b}\right\rangle +1\right\} .\label{21}
\end{align}
It is seen that the two-photon correlation functions are not only related to the variances but also to the correlation functions $\langle X_{i}Y_{i}\rangle$, which are readily evaluated with the use of Eq.~(\ref{13}) and are given by
\begin{align}
\left\langle X_{a}Y_{a}\right\rangle &= \frac{1}{2}i +\frac{1}{2}m\sin2\phi\sinh^{2}\psi ,\label{22}
\end{align}
and
\begin{align}
\left\langle X_{b}Y_{b}\right\rangle &= \frac{1}{2}i + m\sin2\phi\left(1+\frac{1}{2}\sinh^{2}\psi\right) .\label{23}
\end{align}
Then, the two-photon correlation functions obtained from Eq.~(\ref{21}) are
\begin{align}
\left\langle aa\right\rangle = m &-\left[\left(\frac{1}{2}+n\right)g\lambda +m\sin\phi\, e^{-i(\phi+\frac{\pi}{2})}g^{2}\right. \nonumber\\
&\left.  +m\cos\phi\, e^{i\phi}\lambda^{2}\right]\cosh^{2}\psi ,\nonumber\\
\left\langle bb\right\rangle =  me^{2i\phi} &-\left[\left(\frac{1}{2}+n\right)g\lambda +m\sin\phi\, e^{i(\phi-\frac{\pi}{2})}g^{2}\right. \nonumber\\
&\left.  +m\cos\phi\, e^{i\phi}\lambda^{2}\right]\cosh^{2}\psi .\label{24}
\end{align}
The contribution of the first terms on the right-hand side of Eq.~(\ref{24}) obviously correspond to correlations transferred to the modes in the absence of the inter-mode couplings. The second terms correspond to creation of the correlations inside the modes by the inter-mode couplings. It is seen that the effect of the presence of the double coupling between the modes is to create phase-insensitive and phase sensitive contributions to the correlations.  
 
The phase insensitive contributions corresponds to those created when the reservoirs are in thermal states $(m=0)$, and then the correlations are
\begin{align}
\left\langle aa\right\rangle &= \left\langle bb\right\rangle =  -\left(\frac{1}{2}+n\right)g\lambda\cosh^{2}\psi ,\label{25}
\end{align}
This shows that nonzero two-photon correlations arise solely from the presence of the double coupling between the modes. Consequently, uncoupled modes being in thermal states are turned by the double coupling to squeezed states. This is clearly seen when one considers variances of the quadrature components, Eq.~(\ref{17}), which in the case of $m=0$ reduce to
\begin{align}
\langle X^{2}_{a}\rangle &= \langle X^{2}_{b}\rangle = \left(\frac{1}{2}+n\right)\!\left[1 +g(g-\lambda)\!\cosh^{2}\psi\right] ,\nonumber\\
\langle Y^{2}_{a}\rangle &= \langle Y^{2}_{b}\rangle = \left(\frac{1}{2}+n\right)\!\left[1 +g(g+\lambda)\!\cosh^{2}\psi\right] .\label{26}
\end{align}
Evidently, $\langle X^{2}_{i}\rangle < \langle Y^{2}_{i}\rangle ,\ (i=a,b)$ when $\lambda\neq 0$ indicating that the modes are in squeezed states with the variances of $X$ quadratures reduced at the expense of increased variances of $Y$ quadratures. Further inside into the variances reveals that $\langle X^{2}_{i}\rangle$ decreases as $\lambda$ is increasing and a maximum squeezing, i.e. the minimum value of $\langle X^{2}_{i}\rangle$ is reached for $\lambda =g$, i.e. at the exceptional point, in which case $\langle X^{2}_{i}\rangle = (1/2 +n)$. 
Since $\langle X^{2}_{i}\rangle \geq 1/2$, it follows that the maximum reduction of the variances (fluctuations) corresponds to maximally classically squeezed fields.

Conditions for squeezed fluctuations of the modes can be more conveniently examined by investigating the so-called degrees of squeezing defined as
\begin{align}
\eta_{aa} &= \frac{|\left\langle aa\right\rangle|-\left\langle a^{\dag}a\right\rangle}{\left\langle a^{\dag}a\right\rangle} ,\qquad \eta_{bb} = \frac{|\left\langle bb\right\rangle|-\left\langle b^{\dag}b\right\rangle}{\left\langle b^{\dag}b\right\rangle} .\label{27}
\end{align}
Negative values of $\eta_{ii}$, with the limiting value $-1$, indicate classically squeezed field, whereas positive values of $\eta_{ii}$ indicate a quantum squeezed field, and the largest the positive value the greater quantum squeezing.

In order to see it more explicitly it is helpful to introduce normally ordered variances $\langle :X^{2}_{a}:\rangle = \langle X^{2}_{a}\rangle -\frac{1}{2}$, because the condition for quantum squeezing then has the simple form $\langle :X^{2}_{a}:\rangle <0$~\cite{dw83,lk87,zz90,df99,df04,fw14}. In terms of the degrees of the correlations the normally ordered variances are of the forms
\begin{align}
\langle :X^{2}_{a}:\rangle &= \langle a^{\dag}a\rangle\left(2+\eta_{aa}\right) ,\quad \langle :Y^{2}_{a}:\rangle = -\langle a^{\dag}a\rangle\eta_{aa} ,\nonumber\\
\langle :X^{2}_{b}:\rangle &= \langle b^{\dag}b\rangle\left[1+\left(1+\eta_{bb}\right)\cos2\phi\right] ,\nonumber\\
\langle :Y^{2}_{b}:\rangle &= \langle b^{\dag}b\rangle\left[1-\left(1+\eta_{bb}\right)\cos2\phi\right] . \label{27a}
\end{align}
It is clearly seen that in the case of classically squeezed field where $\eta_{ii}\leq 0$ all the normally ordered variances are nonnegative. However, for $\eta_{ii}>0$ the normally ordered variance $\langle :Y^{2}_{a}:\rangle$ is negative and depending on the phase $\phi$ either $\langle :X^{2}_{b}:\rangle$ or $\langle :Y^{2}_{b}:\rangle$ can be negative indicating that the modes are in quantum squeezed states.  

When we specify to thermal reservoirs $(m=0)$ interacting with the modes, we get
\begin{align}
\eta_{aa} &= \eta_{bb} = -\frac{n +\left(\frac{1}{2}+n\right)g\left(g-\lambda\right)\cosh^{2}\psi}{n +\left(\frac{1}{2}+n\right)g^{2}\cosh^{2}\psi} ,\label{28}
\end{align}
which is always negative. Thus, fluctuations of both modes cannot be reduced below the vacuum limit of $\frac{1}{2}$. Therefore, both modes display classically squeezed fluctuations.

Further, we would like to point out that Eq.~(\ref{25}) provides an interesting example of an unexpected result that interactions applied {\it between} the modes create correlations {\it inside} the modes. Such effect is absent when only one type of the interaction, linear or nonlinear, is applied between the modes. 

The phase sensitive contributions to the correlations (\ref{24}) are due to correlations $m$ existing in squeezed reservoirs. A close look at the correlations, Eq.~(\ref{24}), reveals that the two-photon correlations inside the modes are created by the linear and nonlinear coupling processes from the correlations existing in the reservoirs under completely different conditions. It can be seen in Eq.~(\ref{24}) that for $\phi=0$ the phase sensitive contributions are due to the linear coupling process $(\lambda)$, whereas for $\phi=\pi/2$ the contribution to the correlations is done by the nonlinear process $(g)$.

Let us examine how the correlations behave at the exceptional point, when $\lambda =g$. In this case, we obtain from Eq.~(\ref{24}) that
\begin{align}
\left\langle aa\right\rangle &= m -\left(\frac{1}{2}+n+ m\cos2\phi\right)g^{2} ,\nonumber\\
\left\langle bb\right\rangle &=  me^{2i\phi} -\left(\frac{1}{2}+n+m\right)g^{2} ,\label{29}
\end{align}
from which it is apparent that the correlations contained in the mode $a$ are sensitive to phase of the reservoir to which mode $b$ is coupled. Remembering that in the large quantum squeezing limit, $n-m\rightarrow -\frac{1}{2}$, we see that the process of enhancement of the correlations inside the mode $a$ can be significantly reduced and ultimately ceased if one choses $\phi=\pi/2$.  
Thus, for strongly squeezed reservoirs the same dramatic phase dependence exhibited by the population of the mode $a$ also appears in the single-mode two-photon correlations.

\subsection{Correlations between the modes}

Now we turn our attention to the two-mode correlations which could exist between the modes. There could be one-photon and two-photon correlations existing  between the two modes which are determined by the correlation functions $\langle a^{\dag}b\rangle$ and $\langle ab\rangle$, respectively. From Eq.~(\ref{15}), we find that the correlations can be evaluated by calculating  the correlations between quadratures of different modes
\begin{align}
\left\langle a^{\dag}b\right\rangle &= \frac{1}{2}\left[\left\langle X_{a}X_{b}\right\rangle +\left\langle Y_{a}Y_{b}\right\rangle +i\left(\left\langle X_{a}Y_{b}\right\rangle -\left\langle Y_{a}X_{b}\right\rangle\right)\right] ,\label{30}
\end{align}
and
\begin{align}
\left\langle ab\right\rangle &= \frac{1}{2}\left[\left\langle X_{a}X_{b}\right\rangle -\left\langle Y_{a}Y_{b}\right\rangle +i\left(\left\langle X_{a}Y_{b}\right\rangle +\left\langle Y_{a}X_{b}\right\rangle\right)\right] .\label{31}
\end{align}
The correlation functions can be evaluated using Eq.~(\ref{13}) with the help of Eq.~(\ref{05}), and their explicit steady-state expressions are given in the Appendix A. Hence,
by inserting Eqs.~(\ref{A1}) and (\ref{A2}) into Eqs.~(\ref{30}) and (\ref{31}), we obtain
\begin{align}
\langle a^{\dag}b\rangle &= -mg\sin\phi\, e^{i\phi}\cosh^{2}\psi ,\label{32}
\end{align}
and
\begin{align}
\left\langle ab\right\rangle &= -i\left[\left(\frac{1}{2}+n\right)g+m\lambda\cos\phi\, e^{i\phi} \right]\cosh^{2}\psi .\label{33}
\end{align}
From Eq.~(\ref{32}), we see that the interaction of the modes with squeezed reservoirs $(m\neq 0)$ is essential for generating the first-order correlations between the modes that it is not possible to have $\langle a^{\dag}b\rangle\neq 0$ when the modes interact with thermal reservoirs. In addition, the correlations are specifically dependent on the relative phase of the squeezed reservoirs. A comparison of Eqs.~(\ref{32}) and (\ref{33}) shows that the difference that the phase $\phi$ has on the one-photon and two-photon correlations. 
Namely, the choice of phase $\phi=0$ results in $\langle a^{\dag}b\rangle=0$ with simultaneous enhancement of the two-photon correlations $\langle ab\rangle$. Conversely, the choice of phase $\phi=\pi/2$ results in $\langle a^{\dag}b\rangle$ maximal and $\langle ab\rangle$ minimal. Thus, the first-order coherence between the modes can be destroyed or preserved depending on the relative phase of the reservoirs. 

In addition, Eqs.~(\ref{32}) and (\ref{33}), reveal that the one-photon correlations between the modes are created solely by the nonlinear coupling process from the two-photon correlations existing in the squeezed reservoirs. However, the two-photon correlations between the modes are created by both linear and nonlinear coupling processes. The effect of the nonlinear coupling is to create a phase insensitive contribution from the reservoirs thermal noise $(\frac{1}{2}+n)$, whereas the phase sensitive contribution is created by the linear coupling process from the two-photon correlations existing in the squeezed reservoirs. Note that $\langle ab\rangle$ becomes phase insensitive when phase $\phi=\pi/2$.

\begin{figure}[H]
\includegraphics[width=8.0 cm]{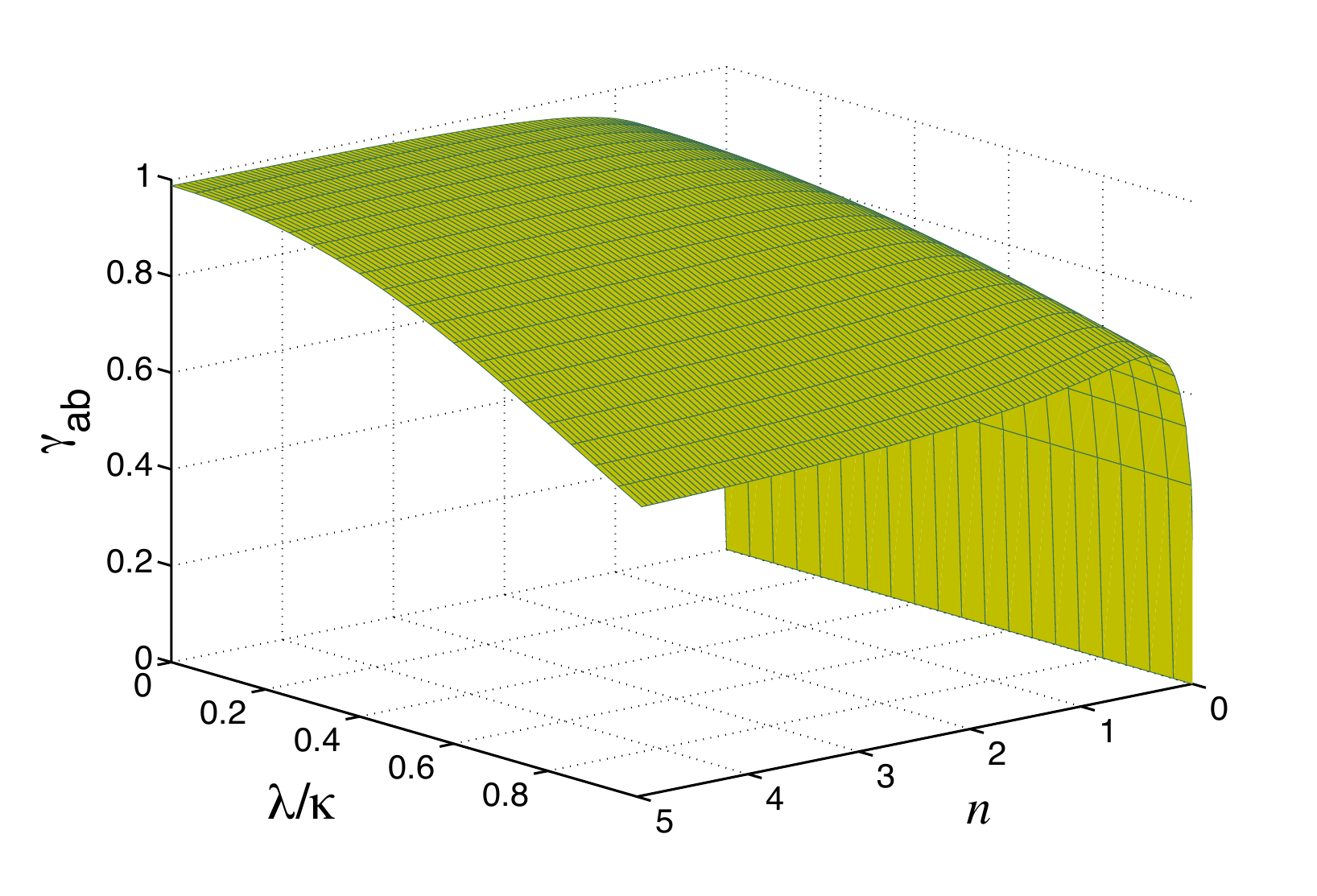}
\caption{Degree of the first-order coherence $\gamma_{ab}$ plotted as a function $n$ and $\lambda/\kappa$ starting from the purely nonlinear coupling $\lambda=0$ to the exceptional point $\lambda=g$, for $m=\sqrt{n(n+1)}$, $\phi=\pi/2$, and $g=0.99\kappa$. \label{fig3}}
\end{figure}

As $\langle a^{\dagger}b\rangle$ can be nonzero, it follows that the modes can be coherent to the first-order, and we can determine the degree of the first-order coherence by considering the quantity $\gamma_{ab} = |\left\langle a^{\dagger}b\right\rangle|/\sqrt{\left\langle a^{\dag}a\right\rangle\left\langle b^{\dag}b\right\rangle}$, which lies between $0$ for mutually incoherent and $1$, for mutually perfectly coherent fields. For the case considered here, the explicit analytical form of $\gamma_{ab}$ can be found using  Eqs.~(\ref{32}), (\ref{18}) and (\ref{19}). With the choice of phase $\phi=\pi/2$, at which $|\langle a^{\dagger}b\rangle|$ is maximal, we obtain
\begin{align}
\gamma_{ab} &= \frac{mg\cosh^{2}\psi}{\sqrt{\left[n +\left(\frac{1}{2}+n\right)g^{2}\cosh^{2}\psi\right]^{2} -\left(mg\lambda\cosh^{2}\psi\right)^{2}}} .\label{34}
\end{align}
Figure~\ref{fig3} shows the coherence function $\gamma_{ab}$ as a function of $n$ and as a function of $\lambda/\kappa$ for $g=0.99\kappa$. It is seen that the degree of coherence increases with $n$ such that for $n>1$ this is close to unity, indicating that the modes are almost perfectly coherent.

In order to distinguish between classical and quantum two-photon correlations present between the modes it is useful to define the degree of mutual two-photon correlations by introducing the quantity $\eta_{ab} = |\left\langle ab\right\rangle|/\sqrt{\left\langle a^{\dag}a\right\rangle\left\langle b^{\dag}b\right\rangle}$, which lies between $0$ and $\infty$. 

However, there is a threshold value of $\eta_{ab}$ at which one can distinguish between the cases of classically and quantum correlated (entangled) modes. In the literature there are several different criteria for detection of entanglement applicable for Gaussian systems~\cite{hh96,rs00,dg00,gm03,sw05,hz06,ad10}. For example, the DGCZ~\cite{dg00} criterion for entanglement of two equally populated modes $(\langle a^{\dag}a\rangle = \langle b^{\dag}b\rangle)$:
\begin{align}
\Delta_{ab} &= \left[\langle \left(X_{b} + Y_{a}\right)^{2}\rangle + \langle\left(Y_{b}+X_{a}\right)^{2}\rangle \right]  \nonumber\\
&= 2 + 4\langle a^{\dag}a\rangle\left(1-\eta_{ab}\right)  ,
\end{align}
clearly shows that the modes are entangled $(\Delta_{ab}<2)$ when $\eta_{ab}>1$.
An another, often used criterion for a quantum (entangle) field, involves the Cauchy-Schwartz inequality~\cite{mw95,gz00}, which for two classically correlated Gaussian modes is of the form
\begin{align}
\chi_{(a,b)} = \frac{(2 + \eta_{aa}^{2})(2 + \eta_{bb}^{2} )}
{\left(1 + \gamma_{ab}^{2} +\eta_{ab}^{2}\right)^{2}}  > 1 .
\end{align}
The inequality is violated for quantum correlated modes, and in the case of $\eta_{aa}=\eta_{bb}$, the inequality is violated when
\begin{align*}
\eta_{ab}^{2} > 1+\eta_{aa}^{2} -\gamma_{ab}^{2} .
\end{align*}
Thus, there is a threshold value of $\eta_{ab}$ which distinguishes between mutually classically and quantum correlated fields. Namely, for classically correlated fields $\eta_{ab}$ can be no larger than $1$, and $\eta_{ab}>1$ can be achieved only for quantum correlated fields.
When the modes interact with squeezed reservoirs, we find that the choices of phase $\phi=0$ and $\phi =\pi/2$ require significantly different conditions for $\eta_{ab}$ to be larger than one. With the help of Eqs.~(\ref{33}), (\ref{18}) and (\ref{19}), we find that
\begin{align}
\eta_{ab} &=  1+\frac{(1-g)\left[\left(\frac{1}{2}+n\right)g +m\lambda\right]\cosh^{2}\psi -n}{n +g\left[\left(\frac{1}{2}+n\right)g+m\lambda\right]\cosh^{2}\psi} ,\label{35}
\end{align}
when $\phi=0$, and 
\begin{align}
\eta_{ab} &= \frac{\left(\frac{1}{2}+n\right)g\cosh^{2}\psi }{\sqrt{\left[n +\left(\frac{1}{2}+n\right)g^{2}\cosh^{2}\psi\right]^{2} -\left(mg\lambda\cosh^{2}\psi\right)^{2}}} .\label{36}
\end{align}
when $\phi=\pi/2$. Since in the exponential amplification regime, the condition for stable steady-state solutions~\cite{mw95,mg67a,mg67b,cg84} requires $g<1$, we see from Eq.~(\ref{35}) that in the case of the ordinary vacuum $(m=n=0)$ it ensures that always $\eta_{ab}$ is larger than one. However, for the thermal or squeezed vacuum, where $n\neq 0$ the first term in the numerator of Eq.~(\ref{35}) may not be large enough to enforce $\eta_{ab}>1$. 

\begin{figure}[H]
\includegraphics[width=8.0 cm]{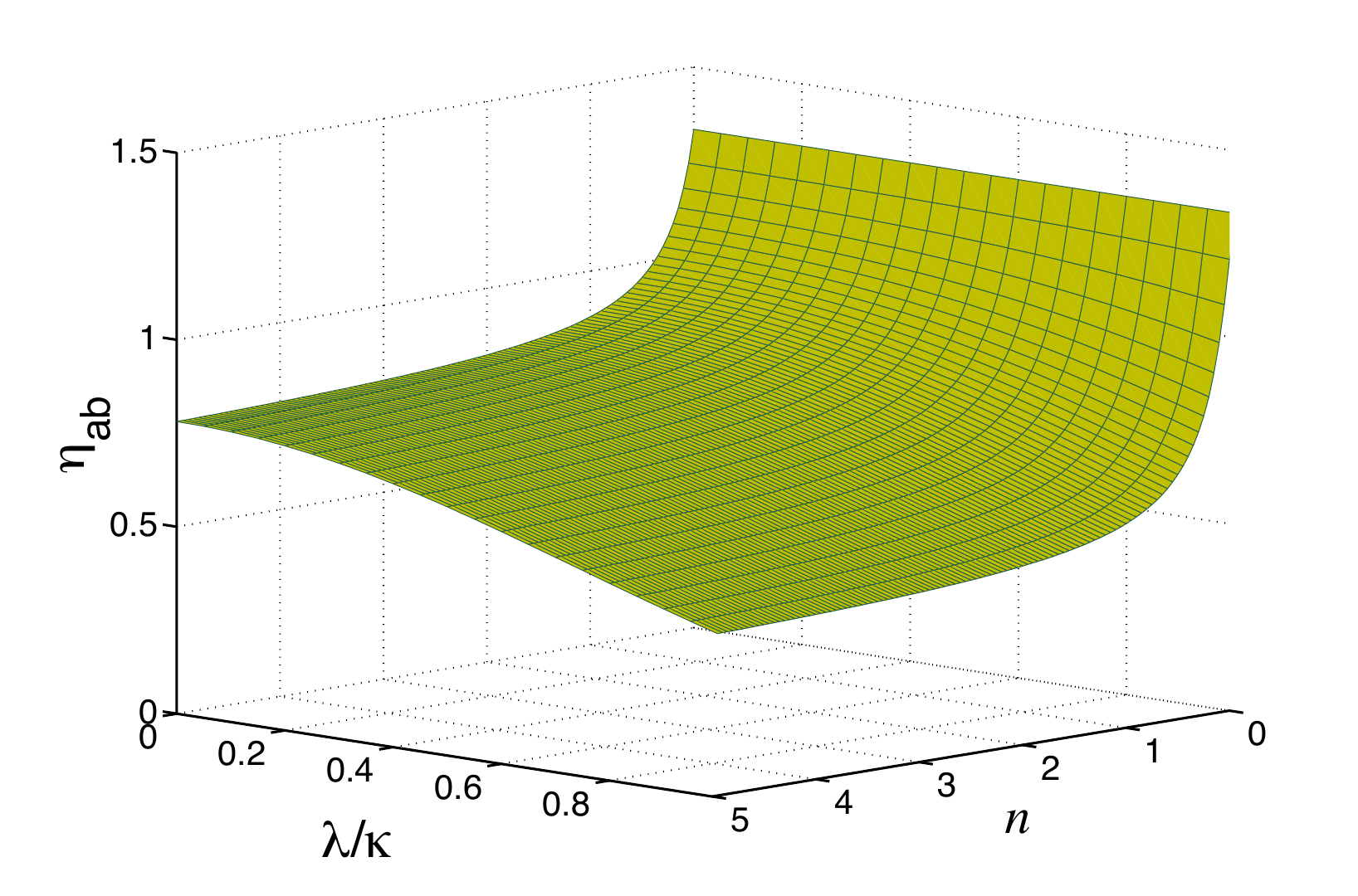}
\caption{Degree of the mutual two-photon correlations $\eta_{ab}$ plotted as a function $n$ and $\lambda/\kappa$ starting from the purely nonlinear coupling $\lambda=0$ to the exceptional point $\lambda=g$, for $m=\sqrt{n(n+1)}$, $\phi=\pi/2$, and $g=0.99\kappa$. \label{fig4}}
\end{figure}

To search for limits imposed by $n$ on $\eta_{ab}$ being larger than one, we plot in  Fig.~\ref{fig4} the degree of two-photon correlations $\eta_{ab}$ as a function of $\lambda$ and as a function of $n$ for $g=0.99$ and $m=\sqrt{n(n+1)}$. The largest value of $\eta_{ab}$ occurs for $n=0$ at which $\eta_{ab}>1$, degrades with an increasing $n$ and turns into values smaller than one for $n\approx 0.5$. The linear interaction has the effect to decrease further the correlations.

A comparison of Fig.~\ref{fig3} and Fig.~\ref{fig4} immediately shows the difference that the number of thermal photons has on the degree of coherence $\gamma_{ab}$ and on the degree of two-photon correlations $\eta_{ab}$. As it is clearly seen, $\eta_{ab}$ is large in the region of $n<0.5$ in which $\gamma_{ab}$ is small, while for the values of $n$ where $\gamma_{ab}$ is large, $\eta_{ab}$ is smaller than one and even tends to zero. For $n\gg1$ the degree of coherence can, in principle, approach one $(\gamma_{ab}=1)$, the perfect coherence.
Thus, unlike the degree of the first-order correlation which approaches maximal values for large $n$, the degree of the two-photon correlations has maximal values for small $n$.
In addition, the perfect coherence $\gamma_{ab}=1$ can be achieved only if the reservoirs are in in quantum squeezed states, $m=\sqrt{n(n+1)}$. For maximally classically squeezed reservoirs, $m=n$, the degree of coherence is significantly less than one.

\section{Oscillatory regime $\lambda>g$.}\label{sec5}

We now turn to consider the case of $\lambda>g$, in which the quadrature components exhibit oscillatory behaviour and, as in the previous section, we study the fluctuation properties of the modes, their populations, single and two-mode correlations. The starting point of our calculations this time is Eq.~(\ref{14}), the solutions for the time evolution of the quadrature components for $\lambda>g$ and we will first consider the steady-state properties of the variances of the quadrature components. 

\subsection{Fluctuation properties of the modes}

Following the same procedure as described in the preceding section, if we evaluate variances of the quadrature components according to Eq.~(\ref{14}), we arrive to the following expressions
\begin{align}
\langle X_{a}^{2}\rangle &= \left(\frac{1}{2}+n+m - m\cos^{2}\phi\sin^{2}\chi\right)\nonumber\\
& -\left(\frac{1}{2}+n-m\cos2\phi\right)g(\lambda-g)\cos^{2}\chi ,\nonumber\\
\langle Y_{a}^{2}\rangle &= \left(\frac{1}{2}+n-m+ m\cos^{2}\phi\sin^{2}\chi\right) \nonumber\\
&+\left(\frac{1}{2}+n+m\cos2\phi\right)g(\lambda+g) \cos^{2}\chi ,
 \nonumber\\
\langle X_{b}^{2}\rangle &= \left(\frac{1}{2}+n+m\cos2\phi -m\cos^{2}\phi\sin^{2}\chi\right)\nonumber\\
& -\left(\frac{1}{2}+n -m\right)g(\lambda -g)\cos^{2}\chi ,\nonumber\\
\langle Y_{b}^{2}\rangle &= \left(\frac{1}{2}+n-m\cos2\phi +m\cos^{2}\phi\sin^{2}\chi\right) \nonumber\\
& +\left(\frac{1}{2}+n +m \right)g(\lambda+g) \cos^{2}\chi ,\label{37}
\end{align}
where $\chi=\arctan(\beta)$. These results are in marked contrast to those found in the exponential region $g>\lambda$ given by Eq.~(\ref{17}). 
In particular, we can see immediately that outside the exceptional point $(g\neq \lambda)$ the variances of the $X$ quadratures are reduced while the variances of the $Y$ quadratures are enhanced by the double coupling. 

The first important fact we can derive from Eq.~(\ref{37}) is that even when the modes interact with thermal reservoirs, the variances of the $X$ quadratures can be reduced below their vacuum value of $1/2$, which indicates that the modes can be found in quantum squeezed states. To show it more explicitly, we set $m=0$ in Eq.~(\ref{37}) and find that
\begin{align}
\langle X_{a}^{2}\rangle &= \langle X_{b}^{2}\rangle =  \left(\frac{1}{2}+n\right)\left[1 - g(\lambda-g)\cos^{2}\chi\right] ,\nonumber\\
\langle Y_{a}^{2}\rangle &= \langle Y_{b}^{2}\rangle = \left(\frac{1}{2}+n\right)\left[1 + g(\lambda+g)\cos^{2}\chi\right]  .\label{38}
\end{align}
Clearly, the fluctuations in $X_{i}$ quadratures can be reduced below $1/2$ if $g\neq \lambda$. This implies that outside the exceptional point the fields of both modes can display quantum squeezed fluctuations.

\begin{figure}[H]
\includegraphics[width=8.0 cm]{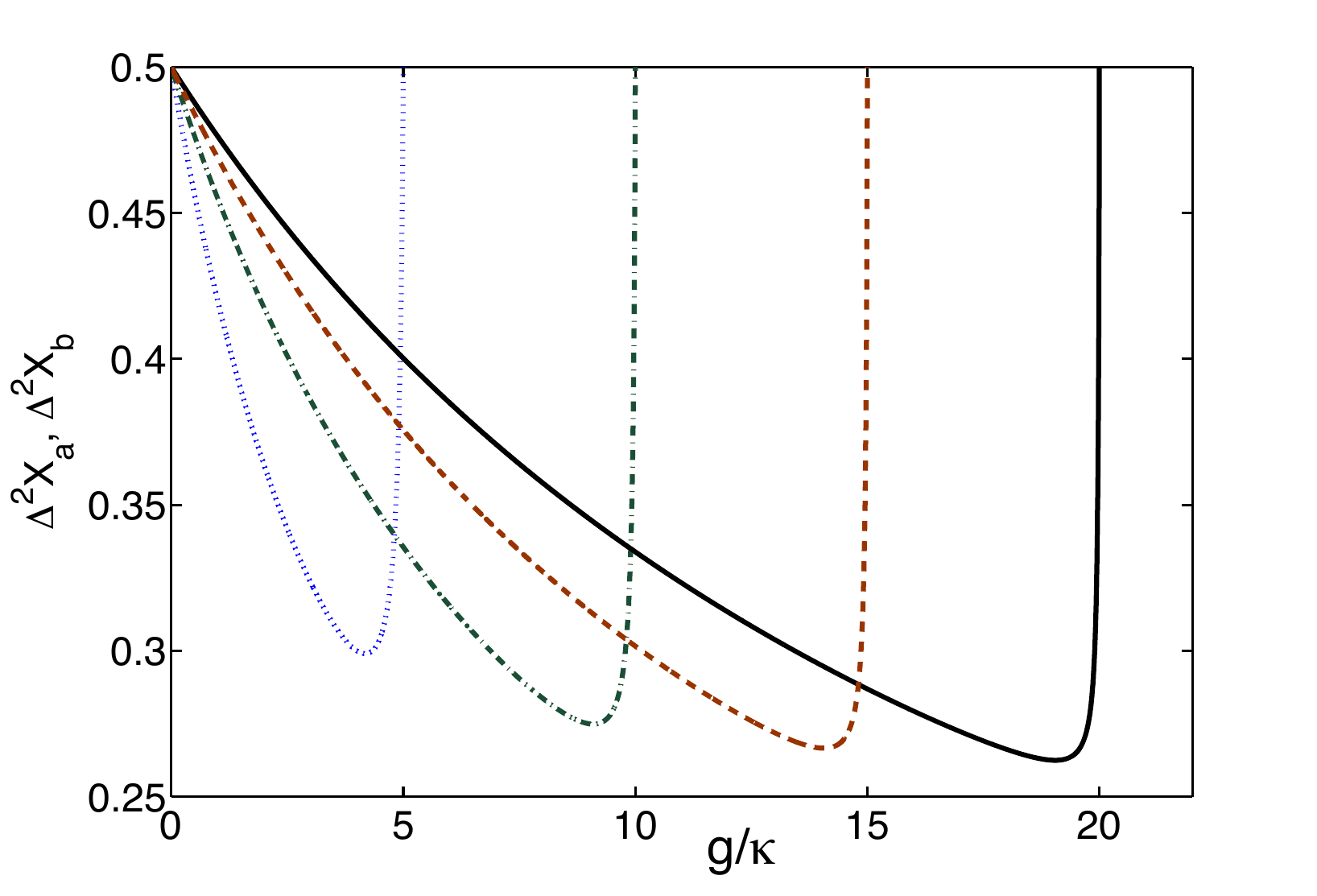}
\caption{Variation of the variances of the $X$ quadratures with $g/\kappa$, starting from the purely linear coupling $g=0$ to the exceptional point $g=\lambda$, for $n=0$ and several different values of $\lambda$: $\lambda =5\kappa$ (blue dotted line), $\lambda=10\kappa$ (green dashed-dotted line), $\lambda=15\kappa$ (red dashed line), $\lambda=20\kappa$ (black solid line).\label{fig5}}

\end{figure}

In Fig.~\ref{fig5} we plot the variances of the $X$ quadratures as a function of the scaled nonlinear coupling strength $g/\kappa$ for $n=0$ and several different values of $\lambda/\kappa$. 
The figure shows clearly that the variances are reduced below $1/2$ for all values of $g<\lambda$ indicating that the modes are in quantum squeezed states and the maximum squeezing occurs for $g$ in the vicinity of the exceptional point. It is seen that approaching the exceptional point cause cessation of quantum squeezing. The largest reduction of the variances is achieved for $\lambda\gg \kappa$, in which case $\Delta^{2}X_{i}\approx 0.25$, so that we may speak of $50\%$ reduction of the fluctuations below the quantum level. Again, we point out that this result is in sharp contrast with the exponential case of $g>\lambda$, Eq.~(\ref{26}), where the variances were reduced to only the vacuum level. 

We may readily adapt the expressions (\ref{37}) for the variances to calculate populations of the modes. 
On making use of variances (\ref{37}) in Eq.~(\ref{16}), we obtain
\begin{align}
\left\langle a^{\dag}a\right\rangle &=  n +\left[\left(\frac{1}{2}+n\right)g^{2} +m\cos2\phi\, g\lambda\right]\cos^{2}\chi ,\label{39}
\end{align}
and
\begin{align}
\left\langle b^{\dag}b\right\rangle &= n +\left[\left(\frac{1}{2}+n\right)g^{2} +m g\lambda\right]\cos^{2}\chi .\label{40}
\end{align}
Apart from the appearance of $\cos\chi$ in place of $\cosh\psi$, Eqs.~(\ref{39}) and (\ref{40}) are formally identical with the corresponding results found in Sec.~\ref{sec4} for the exponential case $g>\lambda$. As for the case $g>\lambda$, only the population of the mode $a$ depends on the phase such that for $\phi=0$ the modes are equally populated and by varying the phase, it is possible to vary the population such that effect of amplification of the population ultimately ceases to occur.

\subsection{Two-photon correlations inside the modes}

We now consider the single-mode two-photon correlations. Using the variances (\ref{37}) in Eq.~(\ref{21}), we find
\begin{align}
\left\langle aa\right\rangle &= m -m\cos\phi\,e^{i\phi}\sin^{2}\chi \nonumber\\
&-\left[\left(\frac{1}{2}+n\right)g\lambda +mg^{2}\cos2\phi\right]\cos^{2}\chi ,\nonumber\\
\left\langle bb\right\rangle &= me^{2i\phi} -m\cos\phi\, e^{i\phi}\sin^{2}\chi \nonumber\\
&-\left[\left(\frac{1}{2}+n\right)g\lambda +mg^{2}\right]\cos^{2}\chi .\label{41}
\end{align}

Further inside into the correlation properties of the modes is gained by considering the degrees of the correlations.
If we restrict ourselves to the case where the doubly coupled modes interact with local thermal reservoirs, we get that the degrees of the correlations are
\begin{align}
\eta_{aa} &= \eta_{bb} = \frac{\left(\frac{1}{2}+n\right)g\left(\lambda-g\right)\cos^{2}\chi - n}{n +\left(\frac{1}{2}+n\right)g^{2}\cos^{2}\chi } .\label{42}
\end{align}
Apparently, the degrees of correlations can have positive values, which is quite different from the exponential amplification case of $g\geq\lambda$, where $\eta_{aa}$ and $\eta_{bb}$ were always negative. It is particularly interesting to note that in the case of the ordinary vacuum state $(n=0)$ of the reservoirs, $\eta_{aa}$ and $\eta_{bb}$ are always positive. Therefore, the fluctuations of the modes can be reduced below the vacuum limit so that the modes can display quantum squeezed fluctuations. The effect of thermal photons $n$ is to diminish the correlations and eventually to turn $\eta_{aa}$ and $\eta_{bb}$ into negative values. In other words, thermal fluctuations can turn the fluctuations of the modes from quantum to classically squeezed fluctuations.

\begin{figure}[H]
\includegraphics[width=8.0 cm]{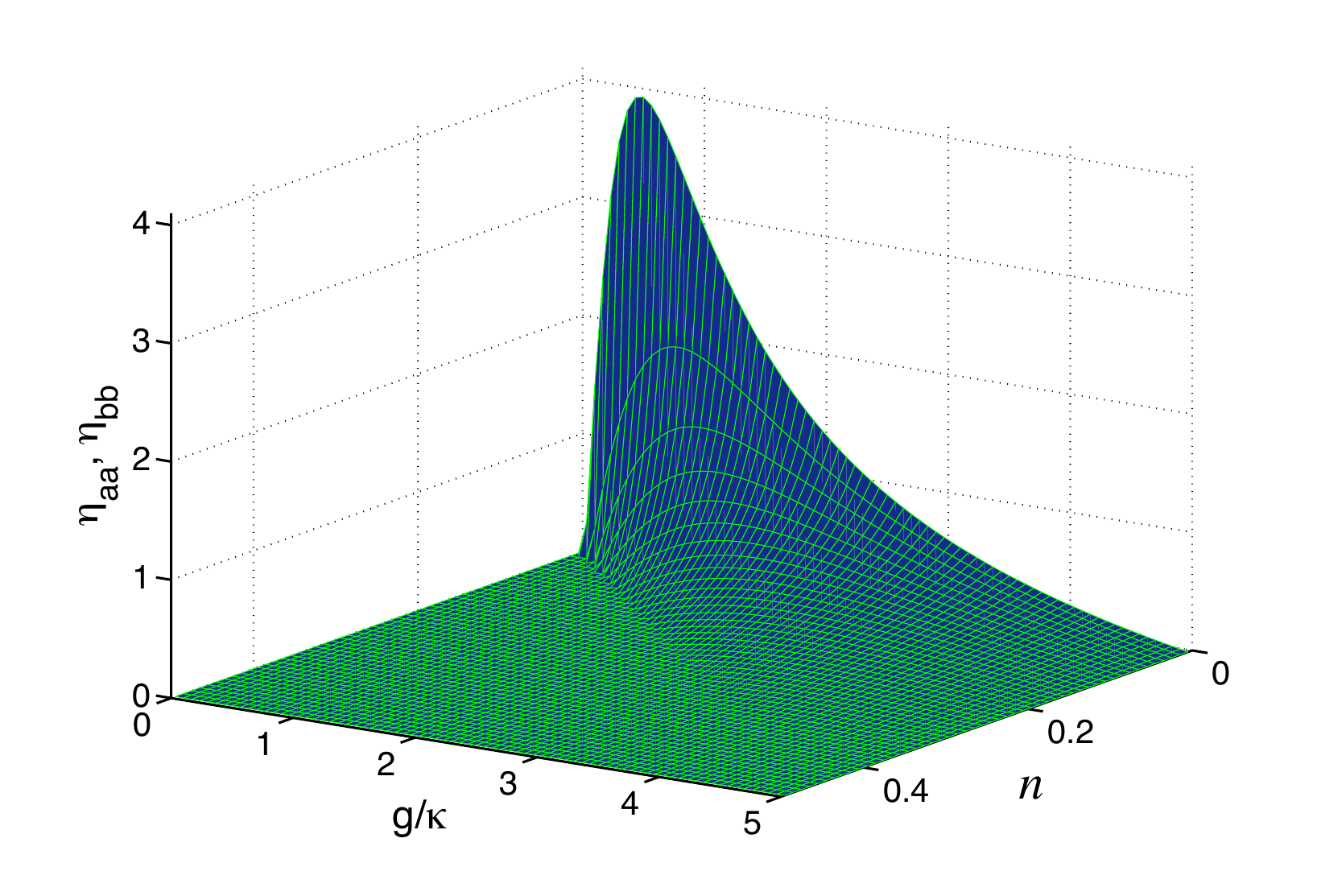}
\caption{Variation of $\eta_{aa}$ and $\eta_{bb}$ with $g/\kappa$ and $n$ for $\lambda=5$ and the modes interacting with thermal reservoirs.\label{fig6}}
\end{figure}

The feature of the correlations just described are easily seen in Fig.~\ref{fig6}, which shows $\eta_{aa}$ and $\eta_{bb}$, as given in Eq.~(\ref{42}) as a function of $g/\kappa$ and $n$ for $\lambda=5\kappa$. The largest value of $\eta_{aa}$ and $\eta_{bb}$ occurs for small $g$ and $n$. It is also seen that with an increasing $g$ the threshold for $\eta_{aa}$ and $\eta_{bb}$ to be positive shifts towards larger $n$.

To continue the analysis of the two-photon correlations, we now assume that the modes interact with squeezed reservoirs. In this case, the degrees of the two-photon correlations depend on the phase $\phi$ such that the choice of the phase $\phi =0$ results in the degrees equal for both modes
\begin{align}
&\eta_{aa} = \eta_{bb} = \frac{\left|m -g\left[\left(\frac{1}{2}+n\right)\lambda +mg\right]\right|\cos^{2}\chi}{n +g\left[\left(\frac{1}{2}+n\right)g +m\lambda\right]\cos^{2}\chi } - 1.\label{43}
\end{align}
The absolute values appearing in the numerator of this equation involve a difference between the two-photon correlations existing in the squeezed reservoirs and correlations generated by the double coupling between the modes. 
\begin{figure}[H]
\includegraphics[width=8.0 cm]{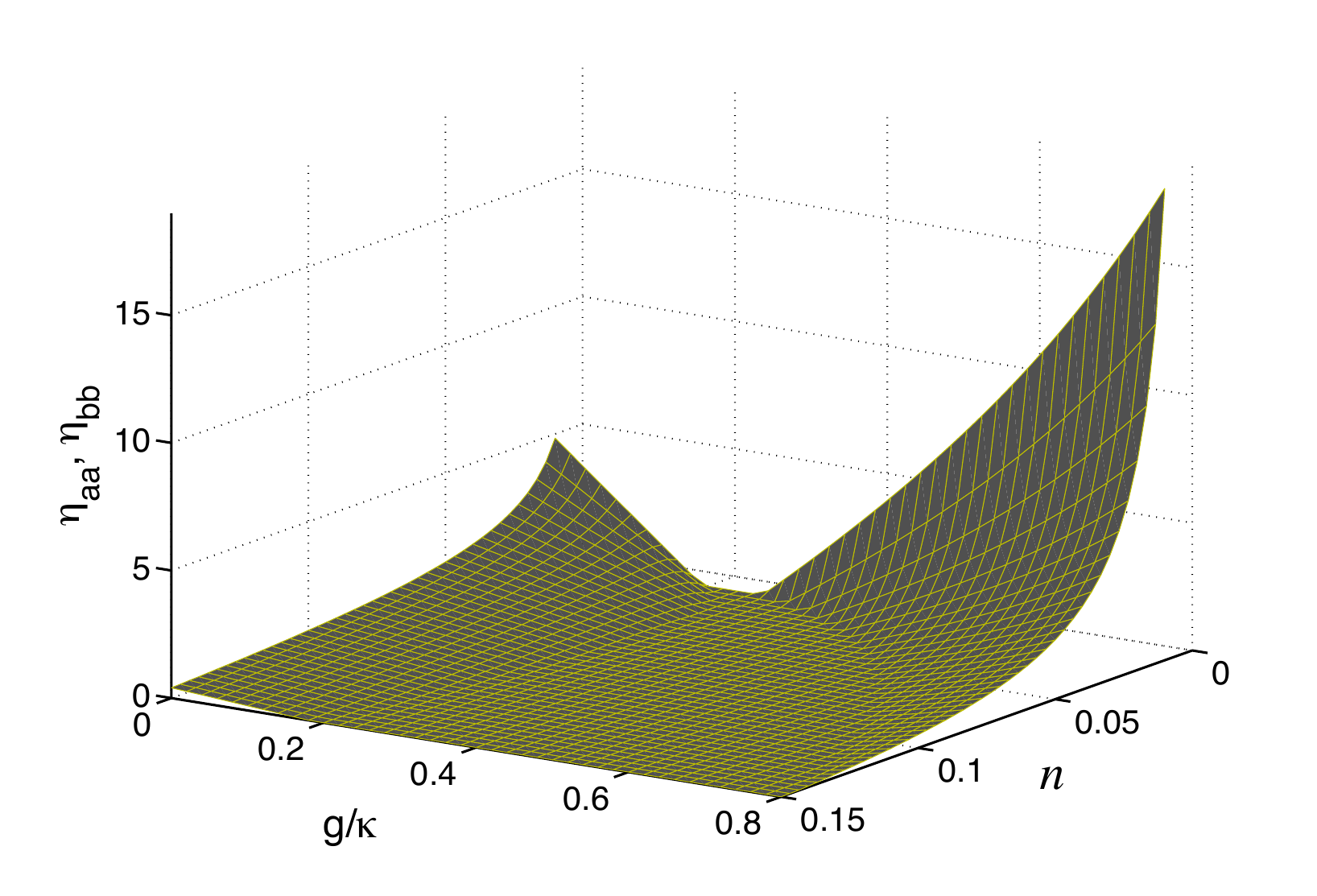}
\caption{Variation of $\eta_{aa}$ and $\eta_{bb}$ with $n$ and $g/\kappa$ starting from the purely linear coupling $g=0$ to the exceptional point $g=\lambda$, for $\lambda=0.8\kappa$ and the modes interacting with quantum squeezed reservoirs, $m=\sqrt{n(n+1)}$. \label{fig7}}
\end{figure}

In Fig.~\ref{fig7}, we plot $\eta_{aa}$ and $\eta_{bb}$ as a function of $g/\kappa$ and $n$ for $\lambda=5\kappa$ and quantum squeezed reservoirs with $m=\sqrt{n(n+1)}$. We can distinguish two separate ranges of the parameters for which the degrees are positive. For the first, occurring for $g<0.2\kappa$ the quantum correlations are those transferred to the modes due to their interaction with the reservoirs. The correlations decrease with an increasing $g$ and ultimately cease at $g=0.2\kappa$. As $g$ increases further quantum correlations appear again showing that the double coupling can generate quantum correlations inside the modes independent of the state of the reservoirs.

\begin{figure}[H]
\includegraphics[width=8.0 cm]{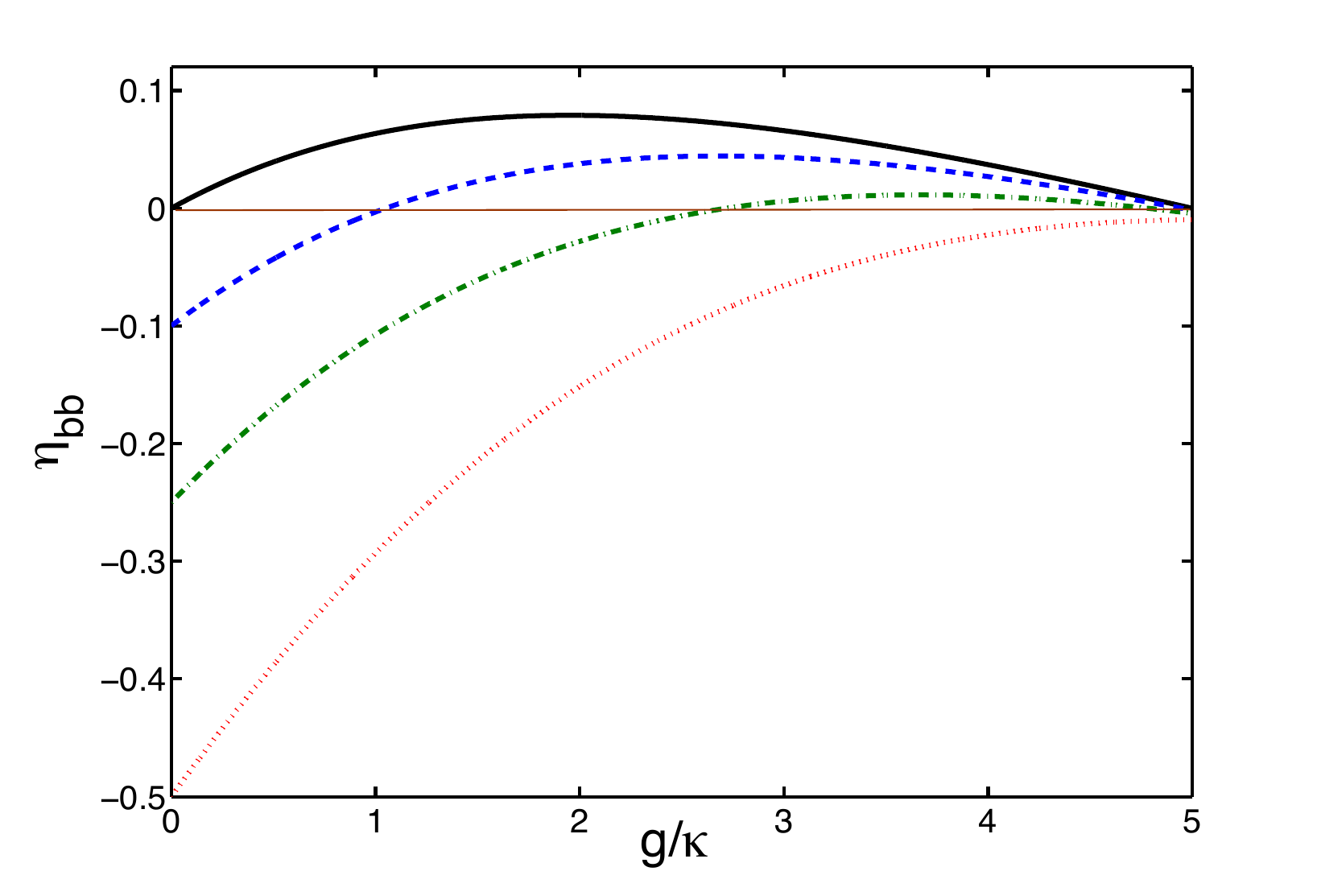}
\caption{Variation of $\eta_{aa}$ and $\eta_{bb}$ with $g/\kappa$ for $n=1$, $\lambda=5\kappa$ and several different values of $m$: $m=0.5n$ (red dotted line), $m=0.75n$ (green dashed-dotted line), $m=0.9n$ (blue dashed line), and $m=n$ (black solid line). \label{fig8}}
\end{figure}

The choice of phase $\phi =\pi/2$ leads to widely different behaviour of the degrees of the correlations of the modes that
\begin{align}
\eta_{aa} &=  \frac{\left|m -g\left[\left(\frac{1}{2}+n\right)\lambda -mg\right]\cos^{2}\chi\right|}{n +g\left[\left(\frac{1}{2}+n\right)g -m \lambda\right]\cos^{2}\chi } - 1,\label{44}
\end{align}
and
\begin{align}
\eta_{bb} &= \frac{m-n+g\left(\frac{1}{2}+n -m\right)(\lambda-g)\cos^{2}\chi}{n +g\left[\left(\frac{1}{2}+n\right)g +m\lambda\right]\cos^{2}\chi} .\label{45}
\end{align}
Since $\lambda>g$, it is clearly seen from the structure of the numerator in Eq.~(\ref{45}) that $\eta_{bb}$ is always positive. Hence, the mode $b$ will exhibit quantum squeezed fluctuations even if the reservoirs are in classically squeezed states. This is illustrated in Fig.~\ref{fig8}, where we plot $\eta_{bb}$ as a function of $g$ for $\lambda=5, n=1$ and several different values of $m\leq n$. For $m=0.5n$, $\eta_{bb}$ is negative over the entire range of $g$ indicating classically squeezed correlations present in the mode. For $m=0.75n$ quantum correlations appear in the restricted range of large $g$. For $m=0.9n$ quantum correlations appear in less restricted range of $g$ and for $m=n$ quantum correlations occur over the entire range of $g$. 

\subsection{Correlations between the modes}

Finally, we consider properties of the two-mode correlations which in the case of $\lambda>g$ follow from Eqs.~(\ref{30}) and (\ref{31}) with the correlations functions involving quadrature operators of different modes given by Eqs.~(\ref{A3}) and (\ref{A4}). Thus, we find that
\begin{align}
\langle a^{\dag}b\rangle &= -\frac{1}{2}gm\sin2\phi \cos^{2}\chi ,\label{46}
\end{align}
and
\begin{align}
\left\langle ab\right\rangle &= -i\!\left[\left(\frac{1}{2}+n+m\sin^{2}\phi\right)\!g+m\cos\phi\, e^{i\phi}\lambda\right]\!\cos^{2}\chi .\label{47}
\end{align}
From the form of the correlations we can see that the one-photon correlation function is different from zero only for choices of phase between $0$ and $\pi/2$ and attains its maximal values for phase $\phi=\pi/4$, at which the linear and nonlinear processes equally contribute to the two-photon correlation function. This result is in marked contrast to that found in the case of $g>\lambda$, where $\langle a^{\dag}b\rangle$ attained a maximal value at $\phi=\pi/2$. At that phase the two-photon correlation function was independent of the correlations $m$. 

\begin{figure}[H]
\includegraphics[width=8.0 cm]{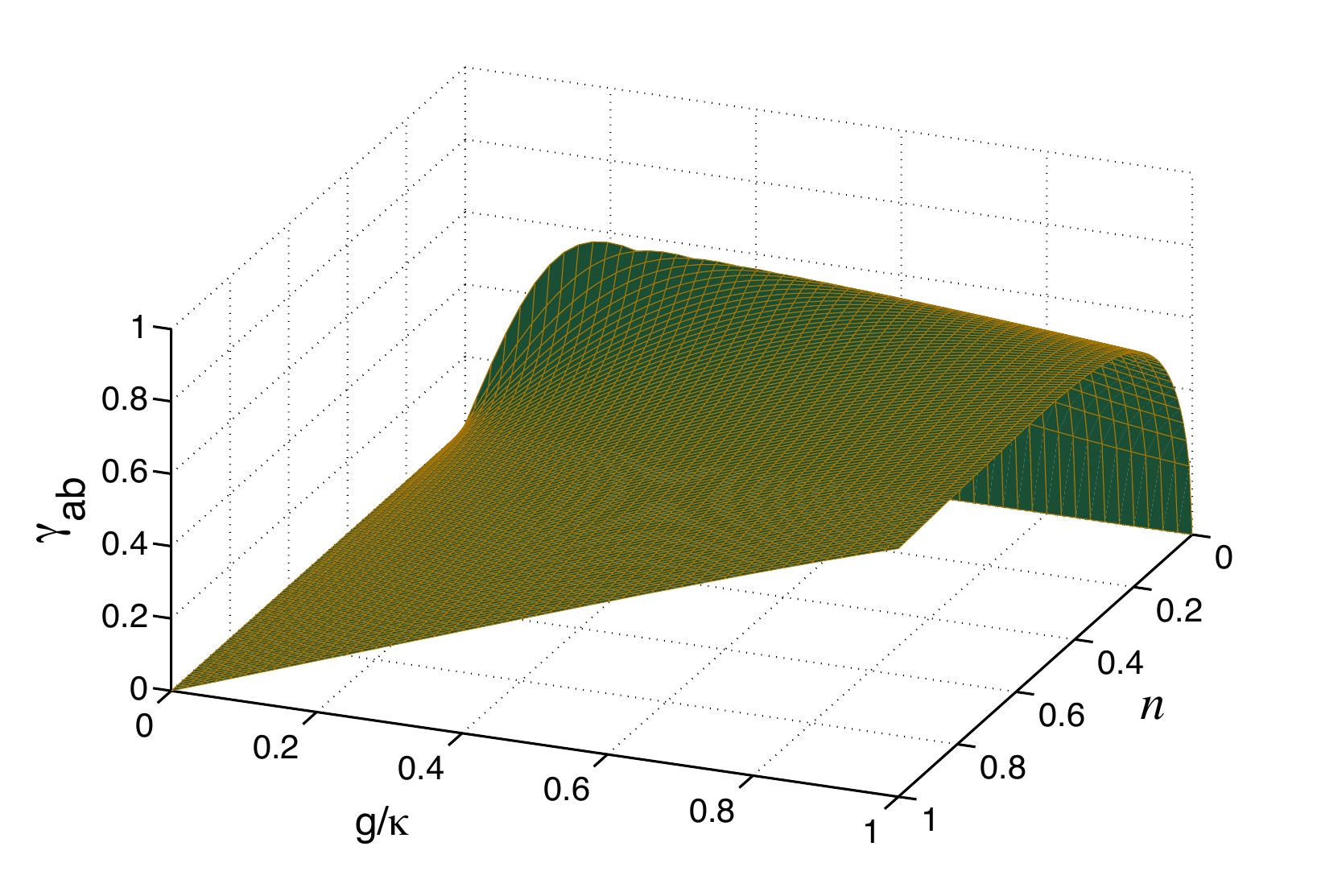}
\caption{Degree of the first-order coherence $\gamma_{ab}$ plotted as a function of $g/\kappa$ and $n$ for $\lambda=\kappa$ and $m=\sqrt{n(n+1)}$.\label{fig9}}
\end{figure}

A better inside into the correlations is obtained by considering the degrees of the correlations. Consider first the degree of the first-order coherence $\gamma_{ab}$. Using Eq.~(\ref{46}) with  the populations given in Eqs.~(\ref{39}) and (\ref{40}), we readily find that for $\phi=\pi/2$ the degree of coherence $\gamma_{ab}$ is
\begin{align}
\gamma_{ab} &= \frac{mg\cos^{2}\chi}{\sqrt{\left[n +\left(\frac{1}{2}+n\right)g^{2}\cos^{2}\chi\right]^{2} -\left(m g\lambda\cos^{2}\chi\right)^{2} }} .\label{48}
\end{align}
The expression is plotted in Fig.~\ref{fig9} for $\lambda=\kappa$ and $m=\sqrt{n(n+1)}$. The degree of the first-order coherence increases with $g$ and attains maximal values which is no larger than $0.6$. This result is significantly different from that obtained in the exponential amplification regime $g>\lambda$, where the degree of the first-order coherence can be as large as $1$. 
To examine properties of the two-mode two-photon correlations, we use Eq.~(\ref{47}), which together with the populations given by Eqs.~(\ref{39}) and (\ref{40}) gives

\begin{align}
\eta_{ab} &=  1+ \frac{(1-g)\left[\left(\frac{1}{2}+n\right)g +m\lambda \right]\cos^{2}\chi -n}{n +g\left[\left(\frac{1}{2}+n\right)g +m\lambda\right]\cos^{2}\chi } ,\label{49}
\end{align}
for the choice of phase $\phi =0$, and
\begin{align}
\eta_{ab} &=  \frac{\left(\frac{1}{2}+n\right)g \cos^{2}\chi}{\sqrt{\left[n +\left(\frac{1}{2}+n\right)g^{2}\cos^{2}\chi\right]^{2} -\left(m g\lambda\cos^{2}\chi\right)^{2} }} ,\label{50}
\end{align}
for the choice of phase $\phi=\pi/2$. 

\begin{figure}[H]
\includegraphics[width=8.0 cm]{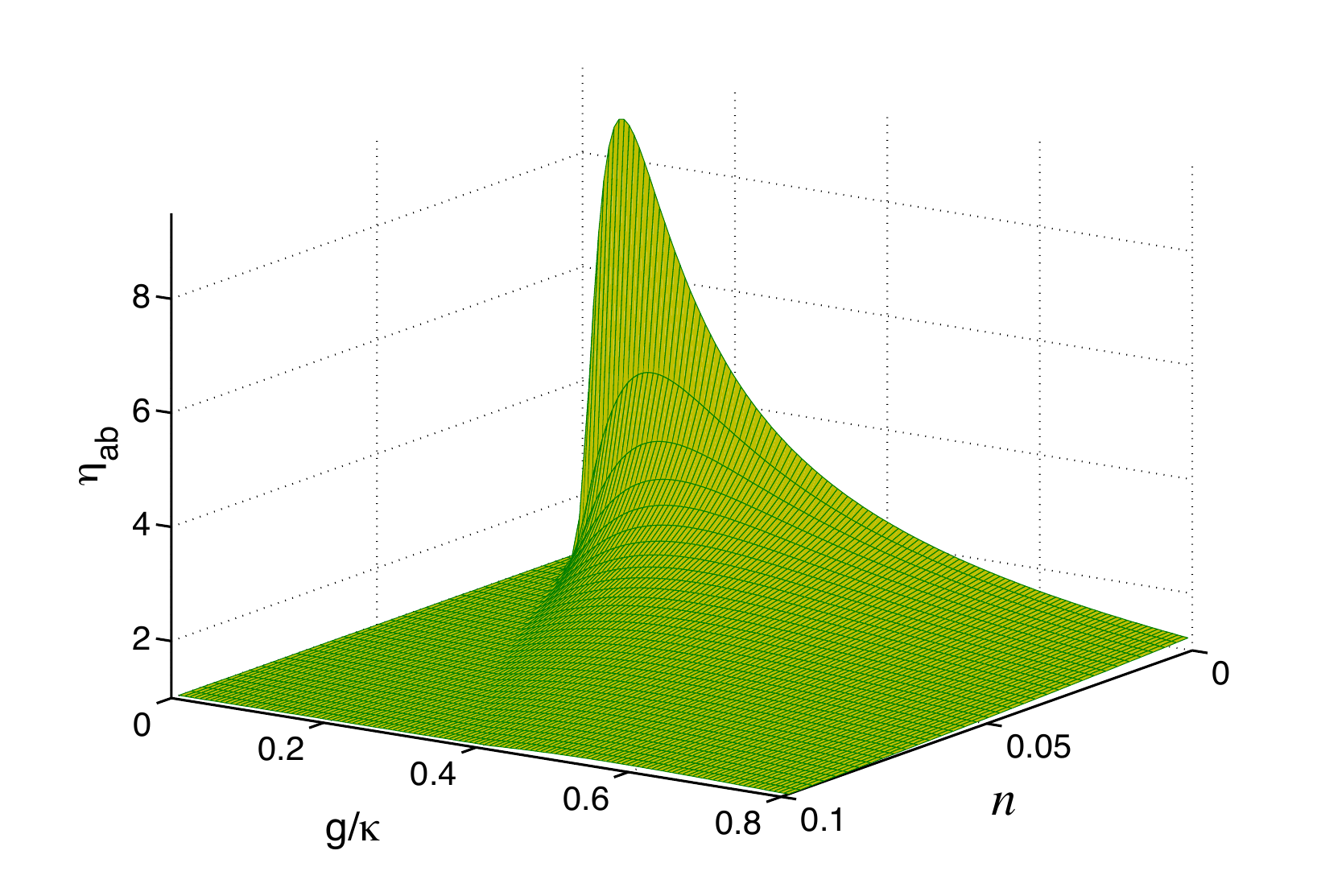}
\caption{Degree of the two-mode two-photon correlations $\eta_{ab}$ plotted as a function of $g/\kappa$ and $n$ for $\lambda=0.8\kappa$ and maximally classically squeezed reservoirs, $m=n$. \label{fig10}}
\end{figure}

It is clear from Eq.~(\ref{49}) that for $\phi=0$ the minimum requirement for $\eta_{ab}$ to be larger than one is that $g$ should be smaller than one. Since in the oscillatory regime there are no limits imposed on $g$, we have that in the case of the ordinary vacuum $(m=n=0)$ the requirement $g<1$ is necessary and sufficient. However, for the thermal or squeezed vacuum it is necessary but not sufficient. For $n\neq 0$ the first term in the numerator of Eq.~(\ref{49}) may not be large enough to enforce $\eta_{ab}>1$. To search for limits imposed by $n$ on $\eta_{ab}$ being larger than one, we plot in  Fig.~\ref{fig10} the degree of two-photon correlations $\eta_{ab}$ as a function of $g$ and as a function of $n$ for $\lambda=0.8$ and $m=n$. The degree of the correlations is initially for $g=0$ equal to one, but increases to values greater than $1$ with an increasing $g$. For $g>0$ and small $n$ the correlations attain maximal values indicating that at the parameter regime strong quantum two-photon correlations exist between the modes.  

\section{Discussion and Conclusions}\label{sec6}

In this paper we have studied the fluctuation and correlation properties of two bosonic modes mutually coupled through a linear photon exchange and nonlinear parametric type processes.
This Hermitian quantum system is known to exhibit non-Hermitian dynamics, which can lead to nonreciprocal (one directional) influences of the modes on each other~\cite{ws23,yt25}. In addition, it results in the appearance of an exceptional point separating two parameter regimes, exponential amplification and oscillatory regimes. The fluctuation and correlation properties of the modes have been investigated by evaluating the stationary expressions for the variances of the quadrature components of the field operators, populations of the modes, and single- and two-mode correlations. We have assumed that apart from the presence of the double coupling, the modes are in contact (interact) with local thermal or squeezed reservoirs, and have shown that the creation of the correlations by the nonreciprocal coupling is strongly influenced by the noise properties of the reservoirs.
The differences in the fluctuations and correlations for these two regimes were studied in details. In particular, for the exponential amplification regime the nonreciprocal coupling tends to convert thermal fluctuations of independent modes being in contact with thermal reservoirs into classically squeezed fluctuations. In the oscillatory regime, however, thermal fluctuations of the modes can be turned by the nonreciprocal coupling into quantum squeezed fluctuations. Apart from the fluctuations, the nonreciprocal coupling can have significant effect on the amplification of the population of the modes. We have shown that when the modes interact with strongly squeezed reservoirs the amplification of the population of one of the modes can be controlled by varying phase of the reservoir interacting with the other mode, and even can be completely ceased.  
We have also discuss conditions under which the modes could be coherent and simultaneously entangled. We have found that the coherence and entanglement exclude each other that strongly coherent modes exhibit classical rather than quantum two-photon correlations, and vice versa, strongly two-photon correlated modes are behaving as being mutually incoherent.

\begin{acknowledgments}
	We would like acknowledge funding by the Minister of Science under the "Regional Excellence Initiative"  program, Project No. RID/SP/0050/2024/1.

\end{acknowledgments}

\appendix

\section{}

In this appendix we give analytic expressions for the stationary correlation functions of quadratures operators of the same and different modes, which are required to evaluate single-mode two-photon correlation functions and two-mode one-photon and two-photon correlation functions.

The required correlation functions are $\langle X_{a}X_{b}\rangle$, $\langle Y_{a}Y_{b}\rangle$, $\langle X_{a}Y_{b}\rangle$, and $\langle Y_{a}X_{b}\rangle$. In the exponential amplification regime, these correlation functions are readily calculated using Eq.~(\ref{13}), and with the help of Eq.~(\ref{05}), we find that 
\begin{align}
\left\langle X_{a}X_{b}\right\rangle + \left\langle Y_{a}Y_{b}\right\rangle  &=  -mg\sin2\phi\, \cosh^{2}\psi ,\nonumber\\
\left\langle X_{a}X_{b}\right\rangle - \left\langle Y_{a}Y_{b}\right\rangle  &=  m\lambda\sin2\phi\, \cosh^{2}\psi ,\label{A1}
\end{align}
and
\begin{align}
&\left\langle X_{a}Y_{b}\right\rangle +  \left\langle Y_{a}X_{b}\right\rangle = -2\!\left[\left(\frac{1}{2}+n\right)\!g+m\lambda\cos^{2}\phi\right]\!\cosh^{2}\psi ,\nonumber\\
&\left\langle X_{a}Y_{b}\right\rangle -  \left\langle Y_{a}X_{b}\right\rangle =  -2mg\sin^{2}\phi \cosh^{2}\psi. \label{A2}
\end{align}

In the oscillatory regime, these correlation functions are calculated using Eq.~(\ref{14}), and with the help of Eq.~(\ref{05}), we find that
\begin{align}
\left\langle X_{a}X_{b}\right\rangle +\left\langle Y_{a}Y_{b}\right\rangle &=  -mg\sin2\phi\cos^{2}\chi ,\nonumber\\
\left\langle X_{a}X_{b}\right\rangle -\left\langle Y_{a}Y_{b}\right\rangle &=  m\lambda \sin2\phi\cos^{2}\chi ,\label{A3}
\end{align}
and
\begin{align}
\left\langle X_{a}Y_{b}\right\rangle +\left\langle Y_{a}X_{b}\right\rangle
&=  -2\!\left[\left(\frac{1}{2}+n\right)\!g +m\lambda \cos^{2}\phi\right]\!\cos^{2}\chi ,\nonumber\\
\left\langle X_{a}Y_{b}\right\rangle -\left\langle Y_{a}X_{b}\right\rangle &= -2mg\sin^{2}\phi \cos^{2}\chi .\label{A4}
\end{align}

 \bibliography{RefJ}

@article{dp18,
	Author = {McDonald, A. and Pereg-Barnea, T. and Clerk, A. A.},
	Journal = {Phys. Rev. X},
	Pages = {041031},
	Title = {Phase-dependent chiral transport and effective non-Hermitian dynamics in a bosonic Kitaev-Majorana chain},
	Volume = {8},
	Year = {2018}}

@article{ps22,
	Author = {del Piero, J. and Slim, J. J. and Verhagen, E.},
	Journal = {Nature},
	Pages = {82},
	Title = {Non-Hermitian chiral phononics through optomechanically induced squeezing},
	Volume = {606},
	Year = {2022}}

@article{ma19,
	Author = {Miri, M.-A. and Alu, A.},
	Journal = {Science},
	Pages = {7709},
	Title = {Exceptional points in optics and photonics},
	Volume = {363},
	Year = {2019}}

@article{or19,
	Author = {Ozdermir, S. K. and Rotter, S. and Nori, F. and Yang, L.},
	Journal = {Nat. Mater.},
	Pages = {783},
	Title = {Parity–time symmetry and exceptional points in photonics},
	Volume = {18},
	Year = {2019}}

@article{wc19,
	Author = {Wang, Y. X. and Clerk, A. A.},
	Journal = {Phys. Rev.},
	Pages = {063834},
	Title = { Non-Hermitian dynamics without dissipation in quantum systems},
	Volume = {99},
	Year = {2019}}

@article{fc20,
	Author = {Flynn, V. P. and Cobanera, E. and Viola, L.},
	Journal = {New J. Phys.},
	Pages = {083004},
	Title = {Deconstructing effective non-Hermitian dynamics in quadratic bosonic Hamiltonians},
	Volume = {22},
	Year = {2020}}

@article{fc21,
	Author = {Flynn, V. P. and Cobanera, E. and Viola, L.},
	Journal = {Phys. Rev. Lett.},
	Pages = {245701},
	Title = {Topology by dissipation: Majorana bosons in metastable quadratic Markovian dynamics},
	Volume = {127},
	Year = {2021}}

@article{ws23,
	Author = {Wanjura, C. C. and Slim, J. J. and del Pino, J. and Brunelli, M. and Verhagen, E. and Nunnenkamp, A.},
	Journal = {Nat. Comm.},
	Pages = {27},
	Title = {Quadrature nonreciprocity: unidirectional bosonic transmission without breaking time-reversal symmetry},
	Volume = {7},
	Year = {2023}}

@article{bb21,
        author  = {Bergholtz, E. J. and Budich, J. C. and Kunst, F. K.},
        title   = {Exceptional topology of non-hermitian systems},
        journal = {Rev. Mod. Phys.}, 
        volume  = {93},
        year    = {2021},
        pages   = {015005}}

@article{df22,
	Author = {Ding, K. and Fang, C. and Ma, G.},
	Journal = {Nat. Rev. Phys.},
	Pages = {745},
	Title = {Non-hermitian topology and exceptional-point geometries},
	Volume = {4},
	Year = {2022}}

@article{pm23,
	Author = {Perina, J. Jr. and Miranowicz, A. and Kalaga, J. K. and Leo\'nski, W.},
	Journal = {Phys. Rev. A},
	Pages = {033512},
	Title = {Unavoidability of nonclassicality loss in PT-symmetric systems},
	Volume = {108},
	Year = {2023}}

@article{am20,
	Author = {Arkhipov, I. I. and Miranowicz, A. and Minganti, F. and Nori, F.},
	Journal = {Phys. Rev. A},
	Pages = {013812},
	Title = {Quantum and semiclassical exceptional points of a linear system of coupled cavities with losses and gain within the Scully-Lamb laser theory},
	Volume = {101},
	Year = {2020}}

@article{he12,
	Author = {Heiss, W. D.},
	Journal = {J. Phys. A: Math. Theor.},
	Pages = {444016},
	Title = {The physics of exceptional points},
	Volume = {45},
	Year = {2012}}

@article{zl19,
	Author = {Zhang, Y. and Lester, B. J. and Gao, Y. Y. and Jiang, L. and Schoelkopf, R. J. and Girvin, S. M.},
	Journal = {Phys. Rev. A},
	Pages = {012314},
	Title = {Engineering bilinear mode coupling in circuit QED: Theory and experiment},
	Volume = {99},
	Year = {2019}}

@article{am24,
	Author = {Ahmadi, B. and Mazurek, P. and Horodecki, P. and Barzanjeh, S.},
	Journal = {Phys. Rev. Lett.},
	Pages = {210402},
	Title = {Nonreciprocal quantum batteries},
	Volume = {132},
	Year = {2024}}

@article{sw24,
	Author = {Slim, J. J. and Wanjura, C. C. and Brunelli, M. and del Pino, J. and Nunnenkamp, A. and Verhagen, E.},
	Journal = {Nature},
	Pages = {767},
	Title = {Optomechanical realization of the bosonic Kitaev chain},
	Volume = {627},
	Year = {2024}}

@article{ba18,
	Author = {Barzanjeh, S. and Aquilina, M. and Xuereb, A.},
	Journal = {Phys. Rev. Lett.},
	Pages = {060601},
	Title = {Manipulating the flow of thermal noise in quantum devices},
	Volume = {120},
	Year = {2018}}

@article{de07,
	Author = {Dimer, F. and Estienne, B. and Parkins, A. S. and Carmichael, H. J.},
	Journal = {Phys. Rev. A},
	Pages = {013804},
	Title = {Proposed realization of the Dicke-model quantum phase transition in an optical cavity QED system},
	Volume = {75},
	Year = {2007}}

@article{rm16,
	Author = {Ruesink, F. and Miri, M. A. and Alu, A. and Verhagen, E.},
	Journal = {Nat. Commun.},
	Pages = {13662},
	Title = {Nonreciprocity and magnetic-free isolation based on optomechanical interactions},
	Volume = {7},
	Year = {2016}}

@article{sh15,
	Author = {Sliwa, K. M. and Hatridge, M. and Narla, A. and Shankar, S. and Frunzio, L. and Schoelkopf, R. J. and Devoret, M. H.},
	Journal = {Phys. Rev. X},
	Pages = {041020},
	Title = {Reconfigurable Josephson circulator/directional amplifier},
	Volume = {5},
	Year = {2015}}

@article{wb20,
	Author = {Wanjura, C. C. and Brunelli, M. and Nunnenkamp, A.},
	Journal = {Nat. Comm.},
	Pages = {3149},
	Title = {Topological framework for directional amplification in driven-dissipative cavity arrays},
	Volume = {11},
	Year = {2020}}

@article{mc20,
	Author = {McDonald, A. and Clerk, A. A.},
	Journal = {Nat. Commun.},
	Pages = {5382},
	Title = {Exponentially-enhanced quantum sensing with non-Hermitian lattice dynamics},
	Volume = {11},
	Year = {2020}}

@article{lz22,
	Author = {Luo, X. W. and Zhang, C. and Du, S.},
	Journal = {Phys. Rev. Lett.},
	Pages = {173602},
	Title = {Quantum squeezing and sensing with pseudo-anti-parity-time symmetry},
	Volume = {128},
	Year = {2022}}

@article{wf09,
	Author = {Wasilewski, W. and Fernholz, T. and Jensen, K. and Madsen, L. S. and Krauter, H. and Muschik, C. and Polzik, E. S.},
	Journal = {Optics Express},
	Pages = {14444},
	Title = {Generation of two-mode squeezed and entangled light in a single temporal and spatial mode},
	Volume = {17},
	Year = {2009}}

@article{vc25,
	Author = {Vimal, V. K. and Cayao, J.},
	Journal = {arXiv:2507.17586},
	Pages = {},
	Title = {Entanglement dynamics in minimal Kitaev chains},
	Volume = {[quant-ph]},
	Year = {2025}}

@article{yt25,
	Author = {Yu, C. and Tian, M. and Kong, N. and Fadel, M. and Huang, X. and He, Q.},
	Journal = {arXiv:2502.04639},
	Pages = {},
	Title = {Exceptional-point-induced nonequilibrium entanglement dynamics in bosonic networks},
	Volume = {[quant-ph]},
	Year = {2025}}

@article{vs18,
	Author = {Vimal, V. K. and Subrahmanyam, V.},
	Journal = {Phys. Rev. A},
	Pages = {052303},
	Title = {Quantum correlations and entanglement in a Kitaev-type spin chain},
	Volume = {98},
	Year = {2018}}

@article{ph16,
	Author = {Peano, V. and Houde, M. and Brendel, C. and Marquardt, F. and Clerk, A. A. },
	Journal = {Nat. Commun.},
	Pages = {10779},
	Title = {Topological phase transitions and chiral inelastic transport induced by the squeezing of light},
	Volume = {7},
	Year = {2016}}

@article{hh11,
	Author = {Hu, Y. C. and Hughes, T. L.},
	Journal = {Phys. Rev. B},
	Pages = {153101},
	Title = {Absence of topological insulator phases in non-Hermitian PT-symmetric Hamiltonians},
	Volume = {84},
	Year = {2011}}

@article{mp15,
	Author = {Malzard, S. and Poli, C. and Schomerus, H.},
	Journal = {Phys. Rev. Lett.},
	Pages = {200402},
	Title = {Topologically protected defect states in open photonic systems with non-Hermitian charge-conjugation and parity-time symmetry},
	Volume = {115},
	Year = {2015}}

@article{ga18,
	Author = {Gong, Z. and Ashida, Y. and Kawabata, K. and Takasan, K. and Higashikawa, S. and Ueda, M.},
	Journal = {Phys. Rev. X},
	Pages = {031079},
	Title = {Topological phases of non-Hermitian systems},
	Volume = {8},
	Year = {2018}}

@article{em18,
	Author = {El-Ganainy, R. and Makris, K. G. and Khajavikhan, M. and Musslimani, Z. H. Rotter, S. and Christodoulides, D. N.},
	Journal = {Nat. Phys.},
	Pages = {11},
	Title = {Non-Hermitian physics and PT symmetry},
	Volume = {14},
	Year = {2018}}

@article{mt18,
	Author = {Malz, D. and Toth, L. D. and Bernier, N. R.},
	Journal = {Phys. Rev. Lett.},
	Pages = {023601},
	Title = {Quantum-limited directional amplifiers with optomechanics},
	Volume = {120},
	Year = {2018}}

@article{bk85,
	Author = {Barnett, S. M. and Knight, P. L.},
	Journal = {J. Opt. Soc. Am. B},
	Pages = {467},
	Title = {Thermofield analysis of squeezing and statistical mixtures in quantum optics},
	Volume = {2},
	Year = {1985}}

@article{mg96,
  title={Generating mutual coherence from incoherence with the help of a phase-conjugate mirror},
  author={Monken, C. H. and Garuccio, A. and Branning, D. and Torgerson, J. R. and Narducci, F. and Mandel, L.},
  journal={Phys. Rev. A},
  volume={53},
  pages={1782},
  year={1996}}

@article{lm98,
	Author = {Mandel, L.},
	Journal = {Pure Appl. Opt.},
	Pages = {927},
	Title = {Anticoherence},
	Volume = {7},
	Year = {1998}}

@book{mw95,
	Author = {Mandel, L. and Wolf, E.},
	Publisher = {Cambridge University Press, Cambridge},
	Title = {Optical coherence and quantum optics},
	Year = {1995}}

@article{sl22,
	Author = {Sun, L. H. and Liu, Y. and Li, Chen. and Zhang, K. K. and Yang, W. X. and Ficek, Z.},
	Journal = {Entropy},
	Pages = {692},
	Title = {Coherence and anticoherence induced by thermal fields},
	Volume = {24},
	Year = {2022}}

@article{mg67a,
	Author = {Mollow, B. R. and Glauber, R. J.},
	Journal = {Phys. Rev.},
	Pages = {1076},
	Title = {Quantum theory of parametric amplification. I},
	Volume = {160},
	Year = {1967}}

@article{mg67b,
	Author = {Mollow, B. R. and Glauber, R. J.},
	Journal = {Phys. Rev.},
	Pages = {1097},
	Title = {Quantum theory of parametric amplification. II},
	Volume = {160},
	Year = {1967}}

@book{df04,
	Author = {Drummond, P. D. and Ficek, Z.(eds)},
	Title = {Quantum squeezing},
	Publisher = {Springer, New York},
	Year = {2004}}

@article{hm15,
	Author = {Heuer, A. and Menzel, R. and Milonni, P.W.},
	Journal = {Phys. Rev. A},
	Pages = {033834},
	Title = {Complementarity in biphoton generation with stimulated or induced coherence},
	Volume = {92},
	Year = {2015}}

@article{mh19,
	Author = {Menzel, R. and Heuer, A. and Milonni, P.W.},
	Journal = {Atoms},
	Pages = {27},
	Title = {Entanglement, complementarity, and vacuum fields in spontaneous parametric down-conversion},
	Volume = {7},
	Year = {2019}}

@article{cg84,
	Author = {Collett, M. J. and Gardiner, C. W},
	Journal = {Phys. Rev. A},
	Pages = {1386},
	Title = {Squeezing of intracavity and traveling-wave light fields produced in parametric amplification},
	Volume = {30},
	Year = {1984}}

@article{lp16,
	Author = {Lahteenmaki, P. and Paraoanu, G. S. and Hassel, J. and Hakonen, P. J.},
	Journal = {Nat. Commun.},
	Pages = {12548},
	Title = {Coherence and multimode correlations from vacuum fluctuations in a microwave superconducting cavity},
	Volume = {7},
	Year = {2016}}

@article{sl12,
	Author = {Sun, L. H. and Li, G.-X. and Ficek, Z.},
	Journal = {Phys. Rev. A.},
	Pages = {022327},
	Title = {First-order coherence versus entanglement in a nanomechanical cavity},
	Volume = {85},
	Year = {2012}}

@article{ga86,
	Author = {Agarwal, G. S.},
	Journal = {Phys. Rev. A},
	Pages = {11584},
	Title = {Anomalous coherence functions of the radiation fields},
	Volume = {33},
	Year = {1986}}

@article{hz06,
	Author = {Hillery, M. and Zubairy, M.S.},
	Journal = {Phys. Rev. Lett.},
	Pages = {050503},
	Title = {Entanglement conditions for two-mode states},
	Volume = {96},
	Year = {2006}}

@article{sw05,
	Author = {Stobi\'nska, M. and W\'odkiewicz, K.},
	Journal = {Phys. Rev. A},
	Pages = {032304},
	Title = {Witnessing entanglement with second-order interference},
	Volume = {71},
	Year = {2005}}

@article{rs00,
	Author = {Simon, R.},
	Journal = {Phys. Rev. Lett.},
	Pages = {2726},
	Title = {Peres-Horodecki separability criterion for continuous variable systems},
	Volume = {84},
	Year = {2000}}

@article{hh96,
	Author = {Horodecki, M. and Horodecki, P. and Horodecki, R.},
	Journal = {Phys. Lett. A},
	Pages = {1},
	Title = {Separability of mixed states: necessary and sufficient conditions},
	Volume = {223},
	Year = {1996}}

@article{ad10,
	Author = {Adesso, G. and Datta, A.},
	Journal = {Phys. Rev. Lett.},
	Pages = {030501},
	Title = {Quantum versus classical correlations in Gaussian states},
	Volume = {105},
	Year = {2010}}

@book{fw14,
	Author = {Ficek, Z. and Wahiddin, M. R.},
	Publisher = {Pan Stanford, Singapore},
	Title = {Quantum optics for beginners},
	Year = {2014}}

@article{pb25,
	Author = {Perina, J. Jr. and Bartkiewicz, K. and Chimczak, G. and Kowalewska-Kudlaszyk, A. and Miranowicz, A. and Kalaga, J. K. and Leo\'nski, W.},
	Journal = {Phys. Rev. A},
	Pages = {043545},
	Title = {Quantumness and its hierarchies in PT-symmetric down-conversion models},
	Volume = {112},
	Year = {2025}}

@article{bx25,
	Author = {Barzanjeh, S. and Xuereb, A. and Alu, A. and Mann, S. A. and Nefedkin, N. and Peano, V. and Rabl, P.},
	Journal = {arXiv:2508.03945},
	Pages = {},
	Title = {Nonreciprocity in quantum technology},
	Volume = {[quant-ph]},
	Year = {2025}}

@article{az23,
	Author = {Akram, J. and Zheng, C.},
	Journal = {Sci. Rep.},
	Pages = {8542},
	Title = {Theoretical investigation of dynamics and concurrence of entangled PT and anti-PT symmetric polarized photons},
	Volume = {13},
	Year = {2023}}

@article{zz24,
	Author = {Zhihang, L. and Zheng, C.},
	Journal = {Symmetry},
	Pages = {584},
	Title = {Non-Hermitian quantum Renyi entropy dynamics in anyonic-PT symmetric systems},
	Volume = {16},
	Year = {2024}}

@article{dg00,
	Author = {Duan, L.M. and Giedke, G. and Cirac, J.I and Zoller, P.},
	Journal = {Phys. Rev. Lett.},
	Pages = {2722},
	Title = {Inseparability criterion for continuous variable systems},
	Volume = {84},
	Year = {2000}}

@article{gm03,
	Author = {Giovanneti, V. and Mancini, S. and Vitali, D. and Tombesi, P.},
	Journal = {Phys. Rev. A},
	Pages = {022320},
	Title = {Characterizing the entanglement of bipartite quantum systems},
	Volume = {67},
	Year = {2003}}

@article{dw83,
	Author = {Walls, D. F.},
	Journal = {Nature},
	Pages = {141},
	Title = {Squeezed states of light},
	Volume = {306},
	Year = {1983}}

@article{lk87,
	Author = {Loudon, R. and Knight, P. L.},
	Journal = {J. Mod. Opt.},
	Pages = {709},
	Title = {Squeezed light},
	Volume = {34},
	Year = {1987}}

@article{zz90,
	Author = {Zaheer, K. and Zubairy, M. S.},
	Journal = {Adv. At. Mol. Opt. Phys.},
	Pages = {143},
	Title = {Squeezed states of the radiation field},
	Volume = {28},
	Year = {1990}}

@book{gz00,
	Author = {Gardiner, C. W. and Zoller, P.},
	Publisher = {Springer, Berlin, New York},
	Title = {Quantum noise},
	Year = {2000}}

@article{df99,
	Author = {Dalton, B. J. and Ficek, Z. and Swain, S.},
	Journal = {J. Mod. Opt.},
	Pages = {379},
	Title = {Atoms in squeezed light fields},
	Volume = {46},
	Year = {1999}}

\end{document}